\documentclass{amsart}
\usepackage{amssymb,stmaryrd}
\usepackage{amsfonts}
\usepackage{amstext}
\usepackage{algorithmic}
\usepackage{algorithm}
\usepackage{graphicx}
\usepackage{epstopdf}
\usepackage[all]{xy}

\numberwithin{equation}{section}
\newtheorem{theorem}{Theorem}[section]

\newtheorem{proposition}[theorem]{Proposition}
\newtheorem{corollary}[theorem]{Corollary}

\theoremstyle{definition}

\newtheorem{case}[theorem]{Case}

\theoremstyle{remark}

\newcommand{\R}{{\mathbb{R}}}

\newcommand{\C}{{\mathbb{C}}}
\newcommand{\Z}{{\mathbb{Z}}}

\newcommand{\<}{{\langle}}

\renewcommand{\>}{{\rangle}}

\newcommand{\tens}{\otimes}

\newcommand{\id}{{\rm id}}

\newcommand{\extd}{{\rm d}}
\newcommand{\del}{{\partial}}

\begin{document}

\title{Gravity induced from quantum spacetime}
\keywords{noncommutative geometry, quantum groups, quantum gravity}

\subjclass[2000]{Primary 81R50, 58B32, 83C57}

\author{Edwin J. Beggs \& Shahn Majid}
\address{Department of Mathematics, University of Swansea\\
Singleton Parc, SA2 8PP, UK\\
\& \\
Queen Mary, University of London\\
School of Mathematics, Mile End Rd, London E1 4NS, UK}

\email{E.J.Beggs@swansea.ac.uk, s.majid@qmul.ac.uk}

\date{Ver4.3: Sept 2013}

\begin{abstract}
We show that tensoriality constraints in noncommutative Riemannian geometry in the $2$-dimensional bicrossproduct model quantum spacetime algebra $[x,t]=\lambda x$ drastically reduce the moduli of possible metrics $g$ up to normalisation to a single real parameter which we interpret as a time in the past from which all timelike geodesics emerge and a corresponding time in the future at which they all converge.  Our analysis also implies a reduction of moduli in $n$-dimensions and we study the suggested spherically symmetric classical geometry in $n=4$ in detail,  identifying two 1-parameter subcases where the Einstein tensor matches that of a perfect fluid for (a) positive pressure, zero density and (b) negative pressure and positive density with ratio $w_Q=-{1\over 2}$. The classical geometry is conformally flat and its geodesics motivate new coordinates which we extend to the quantum case as a new description of the quantum spacetime model as a quadratic algebra. The noncommutative Riemannian geometry is fully solved for $n=2$ and includes the quantum Levi-Civita connection and a second, nonperturbative,  Levi-Civita connection which blows up as $\lambda\to 0$. We also propose a `quantum Einstein tensor' which is identically zero for the main part of the moduli space of connections (as classically in 2D). However, when the quantum Ricci tensor and metric are viewed as deformations of their classical counterparts  there would be an $O(\lambda^2)$ correction to the classical Einstein tensor and an $O(\lambda)$ correction to the classical metric.   \end{abstract}
\maketitle 

\section{Introduction} 

Although the full theory of quantum gravity remains elusive, it seems reasonable that it should in some effective limit recover classical gravity and GR as a parameter $\lambda$, assumed to be the Planck scale, is turned off. If so then we should also be able to step back from this limit as a semiclassical effective description in which classical gravity and GR are subject to Planck scale corrections. It has been argued since the 1980s\cite{Ma:pla} that these corrections should plausibly take the form of noncommutative spacetime (and more generally of noncommutative position-momentum space) in which the notion of geometry is `quantized' so as to allow noncommutative coordinates:
\begin{equation}\label{semi}\left\{{\rm Classical\ geometry}\right\}\Leftarrow \left\{{\rm Quantum\ geometry}\right\}\Leftarrow\left\{{\rm Quantum\ gravity}\right\}.\end{equation}
The motivation behind this is that the corrections being due to quantum effects plausibly {\em should} show up as noncommutativity. Although the full dynamics of quantum gravity is likely to be extremely complex, there should be an effective sector of the theory where an effective classical geometry and the same with quantum corrections can be identified at least at the phase space level. The argument then goes that if phase space is quantum (effective position and space coordinates do not commute) then surely position (and momentum) among themselves would generically should also noncommute. This motivates among other things what is ultimately a {\em hypothesis} of quantum spacetime as a quantum gravity effect.

Another motivation for the quantum spacetime hypothesis is the widely accepted view that the classical continuum should break down at the  Planck scale. Any probe of the classical geometry would not be able to resolve below the  probe particle wavelength. To resolve smaller and smaller scales the probe particle would accordingly need to be more and more massive until it eventually formed a black hole, i.e. destroyed the geometry it was meant to measure. Accordingly, the continuum assumption is intrinsically unverifiable. Moreover, the continuum assumption implies in field theory modes of arbitrarily high momentum and this is problematic. Cut-off at the Planck scale is also problematic and does not for example account for dark energy as vacuum energy (it would lead to the Planck density) -- rather, an actual theory of geometry as we approach the planck scale is needed. Quantum spacetime goes beyond momentum cut-off or discretisation of spacetime as it retains the geometry in a generalised form and yet succeeds in breaking the continuum as a natural fuzziness whereby spacetime coordinates cannot be simultaneously measured due to their mutual noncommutativity. 

Flat quantum spacetimes have been seriously studied on this or similar basis since the early 1990s, particularly after the algebraic techniques that came out of quantum groups\cite{Ma:book}. Arguably the most extensively studied model in the context of quantum gravity is
the Majid-Ruegg bicrossproduct model quantum spacetime\cite{MaRue}
\begin{equation}\label{mink} [x_i,x_j]=0,\quad  [x_i,t]=\lambda x_i,\quad i,j=1,\cdots,n-1\end{equation}
in the case $n=4$. This has the physically testable prediction of a variable speed of light (see \cite{AmeMa}  following up on speculation in \cite{Ame}) and may eventually be confirmed or disproved by data from the Fermi-Glast satellite now in orbit.  Its 3D version is among models that have been identified in various limits of 3D quantum gravity with point sources, see \cite{MaSch} for an overview. This and other convincing models have the merit of the Poincar\'e symmetry of classical flat spacetime being preserved but as a quantum Poincar\'e symmetry. This is both typical and desirable given that the classical Poincar\'e group does not typically act on quantum algebras. The quantum Poincar\'e group for the above model is of a certain bicrossproduct type and was shown in \cite{MaRue} to be isomorphic (but with a different interpretation of the generators) to a quantum `Poincar\'e group' in \cite{Luk} that had been proposed from quantum group theory without an action on an actual quantum spacetime. Note that we use conventions where $\lambda=\imath\lambda_P$ is imaginary and $\lambda_P=\kappa^{-1}$ in older conventions.  The bicrossproduct quantum group construction and (\ref{mink}) for $n=3$ were in \cite{Ma:pla} and associated papers, where the algebra occurs as the Lie algebra of what is now viewed as a nonAbelian momentum group. The `quantum group Fourier transform'\cite{Ma:book} provides the necessary equivalence between a nonAbelian momentum group and noncommutative flat spacetime and this was essential to the variable speed of light prediction in \cite{AmeMa}. It is also essential in other models such as \cite{FreMa} where flat quantum spacetime can be seen to emerge from 3D quantum gravity \cite{FreLiv,tHof}.

All of this success has been, however, on the assumption of the above and similar models being quantum versions of flat spacetime. In the intervening years the current authors and others have been developing a systematic `quantum Riemannian geometry' so as to be able to have models with both quantum spacetime and gravity as a true test of these ideas for quantum gravity and their interplay. In this paper we are now ready to ask the obvious question: {\em are quantum spacetimes such as (\ref{mink}) actually flat space quantum Riemannian geometries?} 

Our answer is summarised in the Summary and Discussion section at the end of the paper, but in brief the answer is a rather surprising `no'. Rather, the innocent looking algebra (\ref{mink}), at least in $n=2$ which we solve in detail, has a unique 1-parameter family of quantum metrics and their classical limit is always curved and never the flat Minkowski one. This classical metric is up to normalisation
\begin{equation}\label{2dmetric} g= b\,x^2\,\extd t^2 +(1+b\,t^2)\extd x^2  -2\, b\, x\, t\, \extd x\, \extd t ,\quad b\ne 0. \end{equation}
The parameter $b$ cannot be 0 if $g$ is nondegenerate.  The metric cannot be flat as the Ricci curvarture is not zero. In fact the Ricci curvature is proportional to $g\over x^2$ so is singular all along the $t$-axis. For the Minkowski signature (where $b<0$) and up to a choice of time orientation we find that all timelike geodesics emerge and end respectively at two points $P_\pm$ on the $t$-axis at $t=\pm 1/\sqrt{-b}$, giving a geometric meaning to the parameter $b$ (see Figure~2(b)). What we are claiming is that all this is forced by even a tiny amount of noncommutativity in the algebra (\ref{mink}) if the classical geometry is to be the limit of a quantum one. In particular, the previously expected flat space Minkowski metric on usual Minkowski space is not quantisable so as to extend to the quantum spacetime (\ref{mink}). In the 4D case the story is more complex but there is again suggested a natural curved background associated to the algebra (\ref{mink}) which we can and do study. For certain parameter values the Einstein tensor corresponds to the stress tensor of a perfect fluid with quintessence parameter $w_Q=-{1\over 2}$. Neither of these results are at the level of entirely physical predictions but they are intended as proof of concept. Certainly the $n=2$ case is a toy model which, however, illustrates what we believe to be an important new phenomenon.

This phenomenon can be viewed as a new type of example of a general rigidity of quantum spacetime: the structure of noncommutative algebra is more rigid than that of the commutative case and geometries which seem perfectly possible or ideas which are perfectly unconnected in the classical limit (the far left in (\ref{semi})) are not possible or are interconnected in the quantum case. A previous example of this was the way that in classical geometry the Laplacian or wave operator  is an independent concept but in quantum geometry it is a unified part of the quantum differential structure\cite{Ma:alm}. Such rigidity phenomena mean that the quantum spacetime hypothesis can have explanatory power even without knowing the full theory of quantum gravity.

The above results take place within a certain paradigm or formalism of quantum Riemannian geometry\cite{Ma:rsph, BegMa4, Ma:alm}; a secondary aim of the paper is to refine some aspects of the formalism itself through the exploration of examples (in our view examples go hand in hand with the development of formalism). Specifically the 2D case can be solved completely in that we can find the quantum Levi-Civita (i.e. quantum metric compatible and quantum torsion free)  connection and its quantum Ricci tensor etc. The  technical innovation here is that the necessary  quantum lift from 2-forms to the quantum version of an antisymmetric tensor is uniquely determined by the required symmetry and reality properties of Ricci. Another remarkable feature is that as well as a unique quantum Levi-Civita connection with classical limit there is a {\em second} unique quantum Levi-Civita connection with no limit as $\lambda\to 0$. This can never be seen in classical geometry but could be of interest as a nonperturbative purely quantum possibility. There was a similar story for the quantum Riemannian geometry of quantum $SU_2$ in \cite{BegMa4}. Also, whereas the classical Einstein equations are empty in 2D (the Einstein tensor always vanishes), this is not the case for these deep-quantum geometries. 

We now introduce two key ingredients in the formalism and hence behind the above rigidity results. The first is the differential structure. A classical manifold means both a `space', reflected in our case in its co-ordinate algebra (typically as a $C^*$-algebra but we do not consider such completions here), and  a `differential structure' on the space, which in our case is done as a specification of a set $\Omega^1$ of 1-forms $\extd x_i,\extd t$ etc., and a map $\extd:A\to \Omega^1$, subject to some widely accepted axioms such as the Leibniz rule. It is not assumed that 1-forms and functions commute, so this is more general than classical differential calculus. In the case of (\ref{mink}) there is only one standard calculus of dimension $n$. It was used, for example, in the variable speed of light prediction \cite{AmeMa} and has commutation relations
\begin{equation}\label{bicrosscalc}[\extd x_i,x_j]=0,\quad  [\extd x_i,t]=0,\quad [x_i,\extd t]=\lambda \extd x_i,\quad [t,\extd t]=\lambda\extd t.\end{equation}
This is the smallest reasonable translation-invariant differential calculus on (\ref{mink}), just as on $\R^n$ there is a standard (in the classical case unique) translation-invariant one. Translation here means with respect to the addition law and in the case of algebras such as (\ref{mink}), which are actually enveloping algebras of Lie algebras (being viewed as a quantum spacetime), translation means with respect to additive Hopf algebra structure. Translation-invariant calculi on this type of Hopf algebra are known, cf. the group algebra version in \cite{Ma:cla}, to be classified by matrix representations of the Lie algebra equipped with a choice of cyclic vector in the representation space.  The Lie algebra (\ref{mink}) is solvable and hence has a natural upper-triangular representation. There is a unique  minimial upper-triangular one\cite{Nun}, which is $n$-dimensional and gives (\ref{bicrosscalc}). Another way to justify the calculus is to note that the smallest quantum-Poincar\'e-invariant calculus is $n+1$-dimensional with an extra non-classical dimension $\theta'$ (tied up with the Laplacian\cite{Ma:alm} as mentioned above) and appears to be unique cf\cite{Sit}. Setting $\theta'=0$ gives (\ref{bicrosscalc}) as the canonical $n$-dimensional projection of this. Our result about the flat metric not being allowed necessarily still applies if we use the calculus with $\theta'$.

The second main ingredient behind our result is the axiomatisation of the quantum metric as an element $g\in \Omega^1\tens_A\Omega^1$, where $A$ is our quantum spacetime algebra and $\Omega^1$ is the set of quantum differential 1-forms (specified by the differential calculus). Note that in classical geometry a tensor $g^{ij}(x)$ has tensor indices given a local basis (over $A$) of $\Omega^1$ and has an $x$-dependence that can be associated equally with $i$ or with $j$. This is something we take for granted in tensor calculus but it means that the tensor is actually an element of a tensor product over $A$, which is what the subscripted $\tens_A$ indicates. We require $g$ to be nondegenerate in the sense of a suitable inverse $(\ ,\ ):\Omega^1\tens_A\Omega^1\to A$ and we require $g$ to be central (to commute with all $a\in A$). This last is a natural requirement in the formalism of noncommutative Riemannian geometry without which contractions via the metric are not well-defined. This is because being central means is equivalent to the inverse metric $(\ ,\ )$ obeying $a(\ ,\ )=(a(\ ),\ )$ and $(\ ,(\ )a)=(\ ,\ )a$ for all $a\in A$ (a bimodule map). This  is something we never have to worry about in classical differential geometry (where functions always commute with forms) and which we therefore take for granted. Without this, contractions such as 
\[ (\id\tens(\ ,\ )):\Omega^1\tens_A\Omega^1\tens_A\Omega^1\to \Omega^1\]
would not be well-defined. In other words, this is a natural extension of  familiar tensoriality properties to take account that $\Omega^1$ is noncommutative. This stronger tensoriality is at the heart of our rigidity result for the quantum metric on (\ref{mink})-(\ref{bicrosscalc}). Although our formalism of quantum Riemannian geometry is very different from that of \cite{Con}, the use of differential forms (differential graded algebras) occurs in all main approaches to noncommutative geometry and after that the specification of $g$ and its desirable properties would seem unavoidable. 

A plan of the paper is as follows. In Section~2 we provide our results on the allowed quantum metrics. In Section~3 we study the understand the classical limit $\lambda\to 0$ of the natural metrics for $n>2$ and fully compute the classical geometry for $n=4$. The rigidity problem is actually more severe in $n>2$ and there is no completely central metric. However, it is natural when model-building to focus on functions of $t,r$ (the latter is the radius) and if we only ask for $g$ to commute with these then metrics are possible, this time a 2-parameter family with standard angular part. Section 3 finds the Einstein tensor and geodesics of these classical metrics while Section~4 shows that a classical change of coordinates motivated by the classical geodesic structure also works nicely in the quantum spacetime case.  Sections~ 5, 6 then cover the full noncommutative Riemannian geometry in the $n=2$ case using the algebraic formalism in \cite{Ma:rsph,BegMa4}. Section~5 first solves the noncommutative model at first order in $\lambda$ as a necessary warm-up. The algebraic methods may be less familiar to readers and we introduce them by working through the details for the model in this first order case. Section~6 then solves the model exactly. The moduli of metric-compatible quantum metrics consists of a conic intersected with a line and we find the two Levi-Civita points where the quantum torsion vanishes, of which one has a classical limit.  The model also suggests a  quantum Einstein tensor which vanishes identically on the conic (just as classically the Einstein tensor in 2D vanishes identically). But both this and the metric have quantum corrections compared to their classical counterparts and the former could be relevant to dark energy. We return to these aspects in the discussion Section~7. 

We equip our coordinate algebra with a complex-linear $*$-involution where $x^i,t,r,\omega_i$ are hermitian (in the sense of self-adjoint under $*$.) This would be relevant when
constructing unitary representations but meanwhile in noncommutative geometry the specification of which elements are `real' in this sense plays the role of working with real-valued functions
in ordinary differential geometry. We require $*$ to extend to  differential forms so as to commute with the exterior derivative $\extd$. This in turn requires that $*$ is a graded-involution with respect to the wedge product of forms.

\section{Moduli of quantum metrics}

We first clarify what we are going to mean by central. In fact the centre of the algebra $A$ defined by relations (\ref{mink}) is easily computed from the relations
\[ [f(x),t]= \lambda\sum_i  x_i{\del\over\del x_i}f,\quad [g(t),x_i]=x_i (g(t-\lambda)-g(t))\]
for functions $f,g$, which relations may in turn be deduced from those stated. It follows that $f(x)$ is central iff it has scaling degree 0, for example rational functions such as $x_1/x_2$ etc will be degree 0. For $g(t)$ to be central we need  that $g$ is periodic in imaginary time. Thus the elements $e^{{2\pi\imath\over\lambda}nt}$ are central. However, these elements exist only as an artefact of the finite difference and have no classical limit as $\lambda\to 0$. They are surely not physical and we will exclude these `periodic null modes' of the finite difference derivative from coefficients of our metric. For example if we limit ourselves in the geometry to rational functions of $t$ then there will be no such `periodic null modes'.  

With this proviso, we think of a general element $f(x,t)$ of $A$ as a normal ordered function of $x_i,t$ with the $t$ to the right. Then $[f,t]=0$ implies and is implied by $f$ being degree 0 under scaling of the $x_i$, and $[f,x_i]=0$ tells us that $f=f(x)$ up to periodic null modes.  So the centre up to such modes is exactly the degree 0 functions of $x$ alone. 

Now consider a metric of the arbitrary form
\[ g=\sum_{i,j}a_{ij}\extd x_i\tens\extd x_j+\sum_i b_i(\extd x_i\tens\extd t+\extd t\tens\extd x_i)+ c\, \extd t\tens\extd t\]
where the coefficients obey $a_{ij}=a_{ji}$ (they are all elements of $A$) and where we have assumed `quantum symmetry' in the form $\wedge(g)=0$. Then using the Leibniz rule, and the relations (\ref{bicrosscalc}), we find (summations understood)
\[ [g,t]=[a_{ij},t]\extd x_i\tens \extd x_j+ ([b_i,t]-\lambda b_i)(\extd x_i\tens \extd t+\extd t\tens\extd x_i)+ ([c,t]-2\lambda c)\extd t\tens\extd t\]
\begin{eqnarray*} [g,x_k]&=&[a_{ij},x_k]\extd x_i\tens\extd x_j -\lambda b_i(\extd x_i\tens\extd x_k+\extd x_k\tens\extd x_i)\\
&&+[b_i,x_k](\extd x_i\tens\extd t+\extd t\tens\extd x_i) -\lambda c (\extd x_k\tens\extd t+\extd t\tens\extd x_k)+[c,x_k]\extd t\tens\extd t
\end{eqnarray*}
If we now use that $\extd x_i,\extd t$ are a basis over $A$ we see that $g$ central amounts to 
\[ [a_{ij},t]=0,\quad \forall i,j,\quad [b_i,t]=\lambda b_i,\quad\forall i,\quad  [c,t]=2\lambda c\]
\[ [a_{ij},x_k]=0,\quad \forall k\ne i,j,\quad  [a_{ik},x_k]=\lambda b_i,\quad\forall i,k\]
\[  [b_i,x_k]=0,\quad \forall k\ne i,\quad [b_k,x_k]=\lambda c,\quad [c,x_k]=0,\quad\forall k.\]

\begin{proposition} When $n>2$ and $\lambda\ne 0$ there are no central quantum-symmetric metrics $g$ up to periodic null mode coefficients, other than the degenerate case  $g_{deg}=\sum_{i,j}a_{ij}\extd x_i\tens\extd x_j$ with $a_{ij}$ of scaling degree 0. 
\end{proposition} 
\proof If $n>2$ we can find $k\ne i$ for any $i$ and hence $[b_i,x_k]=0$ tells us that $b_i$ is a function only of $x$. Then the $[b_k,x_k]$ relation tells us that $[b_k,x_k]=0=\lambda c$ so if $\lambda\ne 0$ we conclude that $c=0$. Similarly for any $i$ we can take $k\ne i$ and $[a_{ii},x_k]=0$ tells us that $a_{ii}$ is a function of $x$ only. Then $[a_{kk},x_k]=0=\lambda b_k$ tells us that $b_k=0$ for all $k$. \endproof

\begin{proposition} \label{metricn2} When $n=2$ and $\lambda\ne 0$ there is, up to an overall normalisation and periodic null modes in the coefficients, a 2-parameter family of central quantum symmetric metrics of the form
\[ g=(t^2+2\beta t +\lambda t+\alpha)\extd x\tens\extd x - x (t+\beta)(\extd x\tens\extd t+\extd t\tens\extd x)+     x^2\extd t\tens\extd t\]
where $\alpha,\beta$ are parameters.  The degenerate cases are 
\[ g_{deg}=(\alpha-2t) \extd x\tens\extd x+x(\extd x\tens\extd t+\extd t\tens\extd x),\quad g_{deg}= \extd x\tens\extd x.\]
\end{proposition}
\proof Writing $a=a_{11}$, $b=b_1$, $c$  for the coefficients and $x=x_1$, the equations above are
\[ [a,t]=[c,x]=0,\quad [c,t]=2\lambda c,\quad [b,t]=\lambda b,\quad [a,x]=2\lambda b,\quad [b,x]=\lambda c.\]
 The equation $[c,x]=0$ tells us that $c=c(x)$ up to periodic null modes, which we are ignoring. In this case $[c,t]=2\lambda c$ becomes $x c'(x)=2c$ hence up to normalisation $c=x^2$. Next let $b=\sum b_n(x)t^n$ say and solve $[b,x]=\sum b_n(x)x ((t-\lambda)^n-t^n)=\lambda x^2=\lambda c$. The $t$-finite difference here can only give a result independent of $t$ if $n=1$. We conclude that $b=-x(t+\beta)$ where $\beta$ is a constant of integration. We check $[b,t]=[-x(t+\beta),t]=-\lambda x(t+\beta)=\lambda b$. We have on equation left $[a,x]=x(a(t-\lambda)-a(t))=-2\lambda x (t+\beta)=2\lambda b$. This requires $a(t-\lambda)-a(t)=-2\lambda (t+\beta)$. This is solved by $a=t^2+(2\beta +\lambda)t+\alpha$ for any constant of integration $\alpha$ and up to periodic null modes. The other option for $c$ is $c=0$. Then $b=b(x)$ by the $[b,x]$ equation. The $[b,t]=\lambda b$ equation then tells us that $xb'(x)=b$ so $b=x$ up to normalisation. In this case the $[a,x]=2\lambda b=2\lambda x$ relation gives $a=-2t+\alpha$ up to periodic null modes. The alternative here is $b=0$ which then implies $a=1$ up to normalisation. \endproof

To clarify the $n=2$ case we introduce central 1-forms
\[ v=x\extd t-t\extd x,\quad v^*= (\extd t) x-(\extd x) t\]
then 
\[ g=v^*\tens v +\lambda(\extd x\tens v-v^*\tens\extd x)-\beta(\extd x\tens v+v^*\tens\extd x)+(\alpha-\lambda(\beta+\lambda))\extd x\tens\extd x\]
is an alternate form of the full metric here. This follows after a lengthy computation using the relations
of the differential algebra. As the 1-forms $\extd x, v, v^*$ are central,  $g$ in this form is manifestly central. The degenerate metrics can also be written in terms of these, thus the first one is
\[ g_{deg}=\extd x\tens v+v^*\tens\extd x+(\alpha+\lambda)\extd x\tens\extd x.\]

We also want our metric $g$ to be `hermitian' in the sense that $g$ is invariant under flip of tensor factors and $*$ on each factor. In $n=2$ this has the effect for the full metric that $\beta$ and $\alpha-\lambda(\beta+\lambda)$ should be real. Finally, it is clear from the form of $g$ stated in Proposition~\ref{metricn2} that we can choose a new variable $t'=t+\beta$ which has the same relations in the differential algebra and which can be used to absorb $\beta$ with a different value of $\alpha$, namely $\alpha'=\alpha-\beta(\beta+\lambda)$. Hence we can set $\beta=0$ in the full metric so that for $n=2$ there is in effect only a 1-real parameter moduli of central metrics here up to normalisation. Similarly in the degenerate metric we need $\alpha+\lambda$ real and can set this to zero by a real translation of $t$. 

Finally, we return to the general $n$ case and use polar coordinates for the bicrossproduct model spacetime\cite{Ma:alm} where we replace $\extd x_i$ by $\omega_i=\sum_j e_{ij}\extd x_j$ where $e_{ij}=\delta_{ij}-{x_ix_j\over r^2}$ is projection to the sphere of constant radius at any point and $r^2=\sum_ix_i^2$. One has $\sum_i x_i\omega_i=0$. The angular part of the metric above is $\omega_i\tens\omega_i$. The polar coordinate relations become
\[ [r,t]=\lambda r ,\quad [{x_i\over r},t]=0\]
for the algebra and
\[ [\omega_i,t]=[\omega_i,r]=[\extd r, t]=[\extd r,r]=0,\quad [r,\extd t]=\lambda\extd r,\quad [t,\extd t]=\lambda\extd t.\]
The relations between 1-forms in the exterior algebra are as classically \cite{Ma:alm}
\[ \{\omega_i,\omega_j\}=\{\omega_i,\extd r\}=\{\omega_i,\extd t\}=\{\extd t,\extd r\}=(\extd r)^2=(\extd t)^2=0. \]

\begin{proposition} \label{dhjsvak}  For $\lambda\ne0$ and all dimensions $n>1$, up to  periodic null modes and translation of the time variable, the `hermitian' quantum-symmetric elements $g\in \Omega^1\tens_A\Omega^1$ with standard angular part and that commute with functions of $r,t$ are of the form 
\[  g=\sum_i \omega_i\tens\omega_i+ a\,\extd r\tens\extd r+ b\, (v^*\tens v + \lambda (\extd r\tens v-v^*\tens\extd r)) \]
for real parameters $a,b$ and $v=r\extd t-t\extd r$. 
\end{proposition} 
\proof This is a reworking of the results above noting that  the $\omega_i$ are already central; their form in the metric is assumed to be fixed and  the remainder is in our 2-dimensional bicrossproduct model spacetime algebra with generators $r,t$ and their differentials. The only difference is that we think geometrically of $r>0$ but this does not affect the algebraic computations. \endproof

We will use our results in the form of Proposition~\ref{dhjsvak} in what follows. For $n=2$ we drop the $\omega_i$ term and regard $r$ as the spatial variable, then this is the general form of the central metric (so only one parameter up to an overall normalisation). For $n>2$ this represents the best we can do in terms of a class of metrics that preserve the spatial rotational symmetry and remain as central as possible.

\section{The classical differential geometry} \label{bnjlilygov}
We would now like to look at the classical geometry given by the metric in Proposition~\ref{dhjsvak} with $n=4$ and setting $\lambda\to 0$. Then
\begin{eqnarray*}
&g=& r^2\,(\extd\theta^2+\sin^2\theta\,\extd\phi^2) +b\,r^2\,\extd t^2 +\extd r^2\,(a+b\,t^2) -\, 2\, \,b\,r\,t\, \extd r\, \extd t   .
\end{eqnarray*}
so that its matrix in the given coordinate order and its inverse (the upstairs metric) are
\begin{eqnarray}\label{gij}
g_{ij}=\left(\begin{array}{cccc}b\,r^2 & -brt & 0 & 0 \\-brt & a+b\,t^2 & 0 & 0 \\0 & 0 & r^2 & 0 \\0 & 0 & 0 & r^2\,\sin^2\theta\end{array}\right),\quad g^{ij}\ =\ \left(\begin{array}{cccc}
 \frac{b t^2+a}{a b r^2} & \frac{t}{a r} & 0 & 0 \\
 \frac{t}{a r} & \frac{1}{a} & 0 & 0 \\
 0 & 0 & \frac{1}{r^2} & 0 \\
 0 & 0 & 0 & \frac{\csc ^2(\theta)}{r^2}
\end{array}
\right)\ .
\end{eqnarray}

The Christoffel symbols for this are computed in the Appendix and from these the Ricci tensor and scalar curvature $S$ easily come out as
\begin{eqnarray}\label{classRicci}
R_{ij}\ =\ \frac1a \left(
\begin{array}{cccc}
 -6 b & \frac{6 b t}{r} & 0 & 0 \\
 \frac{6 b t}{r} & -\frac{2 \left(3 b t^2+a\right)}{r^2} & 0 & 0 \\
 0 & 0 & a-3 & 0 \\
 0 & 0 & 0 & (a-3) \sin ^2(\theta) \\
\end{array}
\right)\ ,\quad S\ =\ \frac{2(a-7)}{a\,r^2}\ .
\end{eqnarray}
We see that we have a curvature singularity at $r=0$ along the $t$-axis, although no scalar curvature if $a=7$.  We also have the Einstein tensor
\begin{eqnarray}\label{Einst}
G_{ij}\ =\ R_{ij}-\tfrac12\, S\,g_{ij} \ =\ \left(
\begin{array}{cccc}
 \left(\frac{1}{a}-1\right) b & \frac{(a-1) b t}{a r} & 0 & 0 \\
 \frac{(a-1) b t}{a r} & \frac{-a^2-b t^2 a+5 a+b t^2}{a r^2} & 0 & 0 \\
 0 & 0 & \frac{4}{a} & 0 \\
 0 & 0 & 0 & \frac{4 \sin ^2(p)}{a} \\
\end{array}
\right)\ .
\end{eqnarray}

\subsection{The interpretation of the stress-energy tensor}

The corresponding upstairs index Einstein tensor is
\begin{eqnarray}\label{Einstup}
G^{ij}\ =\ \left(
\begin{array}{cccc}
 \frac{-a^2-b t^2 a+a+5 b t^2}{a^2 b r^4} & -\frac{(a-5) t}{a^2 r^3} & 0 & 0 \\
 -\frac{(a-5) t}{a^2 r^3} & \frac{5-a}{a^2 r^2} & 0 & 0 \\
 0 & 0 & \frac{4}{a r^4} & 0 \\
 0 & 0 & 0 & \frac{4 \csc ^2(\theta)}{a r^4} \\
\end{array}
\right)
\end{eqnarray}
and we recall Einstein's equation 
\begin{eqnarray}
G^{ij}\ ={8\,\pi\,G}\, T^{ij}\ ,
\end{eqnarray}
where $G$ is the gravitational constant, $T^{ij}$ is the stress-energy tensor. We work in units where the speed of light is 1 and consider the energy-momentum tensor of a perfect fluid
(see \cite{WeinCosmology}), which is
\begin{eqnarray}
T^{ij}\ =\ p\,g^{ij}+(p+\rho)\,u^i\,u^j\ .
\end{eqnarray}
Here $u$ is the normalised 4-velocity of the fluid (i.e.\ $g_{ij}\,u^i\,u^j=-1$ as we have spacelike coordinates with metric sign $+1$), $p$ is the pressure, and $\rho$ is the energy density. If the energy-momentum tensor has this form, then we need $G^{ij}-s\,g^{ij}$ to be a degenerate matrix (determinant zero), and this gives three choices for $s$:
\begin{eqnarray} \label{kjgacsvgjc}
&&s\,=\, \frac{4}{a\,r^2}\ ,\quad G^{ij}-s\,g^{ij}\,=\, \left(
\begin{array}{cccc}
 -\frac{a^2+b t^2 a+3 a-b t^2}{a^2 b r^4} & -\frac{(a-1) t}{a^2 r^3} & 0 & 0 \\
 -\frac{(a-1) t}{a^2 r^3} & -\frac{a-1}{a^2 r^2} & 0 & 0 \\
 0 & 0 & 0 & 0 \\
 0 & 0 & 0 & 0 \\
\end{array}
\right) \cr
&& s\,=\,\frac{5-a}{a\,r^2}\ ,\quad G^{ij}-s\,g^{ij}\,=\, \left(
\begin{array}{cccc}
 -\frac{4}{a b r^4} & 0 & 0 & 0 \\
 0 & 0 & 0 & 0 \\
 0 & 0 & \frac{a-1}{a r^4} & 0 \\
 0 & 0 & 0 & \frac{(a-1) \csc ^2(\theta)}{a r^4} \\
\end{array}
\right)  \cr
&&s\,=\, \frac{1-a}{a\,r^2}\ ,\quad G^{ij}-s\,g^{ij}\,=\,\left(
\begin{array}{cccc}
 \frac{4 t^2}{a^2 r^4} & \frac{4 t}{a^2 r^3} & 0 & 0 \\
 \frac{4 t}{a^2 r^3} & \frac{4}{a^2 r^2} & 0 & 0 \\
 0 & 0 & \frac{a+3}{a r^4} & 0 \\
 0 & 0 & 0 & \frac{(a+3) \csc ^2(\theta)}{a r^4} \\
\end{array}
\right)
\end{eqnarray}
A quick look at the second and third cases of (\ref{kjgacsvgjc}) shows that the matrix 
$G^{ij}-s\,g^{ij}$ is not of rank one (i.e.\ the product of a column and row vector) unless $a=1$ (for the second case) or $a=-3$ (for the third case). This means that the second and third cases for a rank one matrix are special cases of the first case. But considering the first case, the matrix $G^{ij}-s\,g^{ij}$ is of rank one only when $a=1$ or $a=-3$. For the sign of $b$, remember that $a\,b$ is negative for $g_{ij}$ to have signature $-+++$ (take the determinant of $g_{ij}$ to see this). 

Accordingly, we have found two cases where $G$ matches a perfect fluid:

\begin{case} \label{vhjkakjg}
We take $a=1$, in which case $b=-\beta^2$ for some real $\beta$. If we set $u=(1/(\beta r),0,0,0)$, then 
$g_{ij}\,u^i\,u^j=-1$ and 
\begin{eqnarray*}
G^{ij}\ =\ \frac{4}{r^2}\,g^{ij}+\frac{4}{r^2}\,u^i\,u^j,\quad p={1\over 2\,\pi\,G\,r^2},\quad \rho=0.
\end{eqnarray*}
\end{case}

\begin{case}
We take $a=-3$, in which case $b=\beta^2$ for some real $\beta$. If we set $u=(t/r,1,0,0)/\sqrt{3}$, then 
$g_{ij}\,u^i\,u^j=-1$ and 
\begin{eqnarray*}
G^{ij}\ =\ -\,\frac{4}{3\,r^2}\,g^{ij}+\frac{4}{3\,r^2}\,u^i\,u^j,\quad p=-{1\over 6\,\pi\,G\,r^2},\quad \rho={1\over 3\,\pi\,G\,r^2};\quad w_Q=-{1\over 2}
\end{eqnarray*}
\end{case}

In the cosmology literature the ratio $w_Q= {p\over \rho}$ is the `quintessence parameter' and has been associated with models of non-constant
cosmological term where, however, this ratio is spatially constant (but allowed to evolve in FRW time). The case $w_Q=-1$ is obeyed by standard dark energy while $w_Q=-{1\over 2}$ is in the middle of the range $-1<w_Q<0$ referred to in \cite{CDS}. However, our background is not exactly FRW type so we are not proposing a direct comparison with standard cosmology. 

\subsection{Geodesic motion}
From the form of the standard geodesic equation
\begin{eqnarray}
\ddot x^a\ =\ -\,\Gamma^a_{bc}\,\dot x^b\,\dot x^c
\end{eqnarray}
with respect to an affine parameter $\tau$, and from the Christoffel symbols in the Appendix, one can see that we have motion confined to a plane $\theta=\pi/2$ (say). For the $\phi$ motion we have
\begin{eqnarray}
\ddot\phi\ =\ -\,2\,\dot r\,\dot\phi/r\ ,
\end{eqnarray}
which gives the usual conservation of angular momentum $\dot\phi\,r^2=K$, a constant. The $r$ equation is
\begin{eqnarray}\label{ddot r}
a\,\ddot r &=& r\,\dot\phi^2+2\,b\,(r\,\dot t^2-2\,t\,\dot r\,\dot t+t^2\,\dot r^2/r) \cr
&=&r\,\dot\phi^2+2\,b\,(r\,\dot t-t\,\dot r)^2/r \ .
\end{eqnarray}
Similarly, the $t$ equation is 
\begin{eqnarray}
a\, \ddot t&=& t\, \dot\phi^2- {2a\dot r\over r^2}(r\, \dot t- t\, \dot r)+ {2bt\over r^2}(r\, \dot t- t\, \dot r)^2\ .\end{eqnarray}
From these equationswe find
\begin{eqnarray} \label{badjosb}
\frac{\extd(r\,\dot t-t\,\dot r)}{\extd\tau} &=& r\,\ddot t-t\,\ddot r= -{2 \dot r\over r}\,(r\,\dot t-t\,\dot r)\ .
\end{eqnarray}
If we set $f=r\,\dot t-t\,\dot r$, then 
\begin{eqnarray*}
0 &=& \frac{\extd\,\log(f)}{\extd\tau} + 2\, \frac{\extd\,\log(r)}{\extd\tau}= \frac{\extd\,\log(r^2\,f)}{\extd\tau}\ 
\end{eqnarray*}
which implies $f=M/r^2$, where $M$ is a constant of motion.  We also have
\begin{eqnarray*}
\frac{f}{r^2} &=& \frac{\dot t}{r}-\frac{t\,\dot r}{r^2}\ =\ \frac{\extd}{\extd\,\tau}\Big(
\frac{ t}{r}\Big)\ =\ \frac{M}{r^4}\ ,
\end{eqnarray*}
so we get
\begin{eqnarray} \label{bxhajv}
 t\ =\ r\Big(\int \frac{M}{r^4}\,\extd\,\tau+c\Big)\ ,
\end{eqnarray}
where $c$ is a constant of integration.

The length squared of the velocity (with respect to proper time) is
\begin{eqnarray*}
r^2\,\dot\phi^2+a\,\dot r^2+b\,(r\,\dot t-t\,\dot r)^2 
&=& \frac{K^2}{r^2}+a\,\dot r^2+\frac{b\,M^2}{r^4}\ .
\end{eqnarray*}
We then have the equations of motion
\begin{eqnarray} \label{jhgscvghj}
\dot r^2\ =\ \frac{s}{a}-\frac{b\,M^2}{a\,r^4}-\frac{K^2}{a\,r^2}\ ,
\end{eqnarray}
where $s=0$ for null geodesics, $s=-1$ for timelike and $s=1$ for spacelike. This means that for timelike curves $\tau$ becomes the proper time. 
Note that the middle term in the right hand side of (\ref{jhgscvghj}) 
is always positive as $a$ and $b$ are of opposite signs. This means that it is always possible to have $\dot r^2\ge0$ for $r$ sufficiently small. Then we have the integral 
\begin{eqnarray}\label{tau}
\tau\ =\pm \int \frac{\extd r}{\sqrt{ \frac{s}{a}-\frac{b\,M^2}{a\,r^4}-\frac{K^2}{a\,r^2} }}\ 
\end{eqnarray}
where the branch of the square root is determined by initial conditions. Now we can rewrite the formula (\ref{bxhajv}) for $t$ 
\begin{eqnarray} \label{trgeodesic}
 t &=& r\Big(\int \frac{M}{r^4}\,\frac{\extd\,\tau}{\extd\,r}\,\extd r+c\Big)=\pm r\Big(\int \frac{M\,\extd r}{r^4\,\sqrt{ \frac{s}{a}-\frac{b\,M^2}{a\,r^4}-\frac{K^2}{a\,r^2} }}+c\Big)
\end{eqnarray}
We also have
\begin{eqnarray*}
\frac{\extd\,\phi}{\extd\,\tau} \ =\  \frac{\extd\,\phi}{\extd\,r}\, \frac{\extd\,r}{\extd\,\tau}  \ =\  \frac{K}{r^2}\ .
\end{eqnarray*}
so we get the integral
\begin{eqnarray} \label{jgvghjc}
\phi\ =\ \phi_0\pm \int \frac{K\,\extd r}{r^2\,\sqrt{ \frac{s}{a}-\frac{b\,M^2}{a\,r^4}-\frac{K^2}{a\,r^2} }}
\end{eqnarray}
For the moment we take the positive branch of the square root. 

Case 1 : Null geodesics, $a=\alpha^2>0$, $b=-\beta^2<0$. We can solve the integral
(\ref{jgvghjc}) with $s=0$ to get (setting $\phi_0=0$)
\begin{eqnarray} \label{kjhgcvxfggs}
r &=& \frac{M\,\beta}{K}\,\sin(\phi/\alpha)\ ,\quad t/r = -\,\frac{\alpha\,K\,\cot(\phi/\alpha)}{\beta^2\,M}+c\  .
\end{eqnarray}
In the case $\alpha=\beta=1=K=M$ we get null geodesics from $r=0,t=-1$ to
$r=0,t=1$ which describe a circle when projected to the $x,y$ plane. This is shown in Figure~1 with the $t$-axis along the longest side of the bounding box. There are six different geodesics shown, with $c=0,\frac25,\frac45,1,\frac32,2$ as we move from the leftmost to the rightmost path.

\begin{figure}
\includegraphics[scale=0.6]{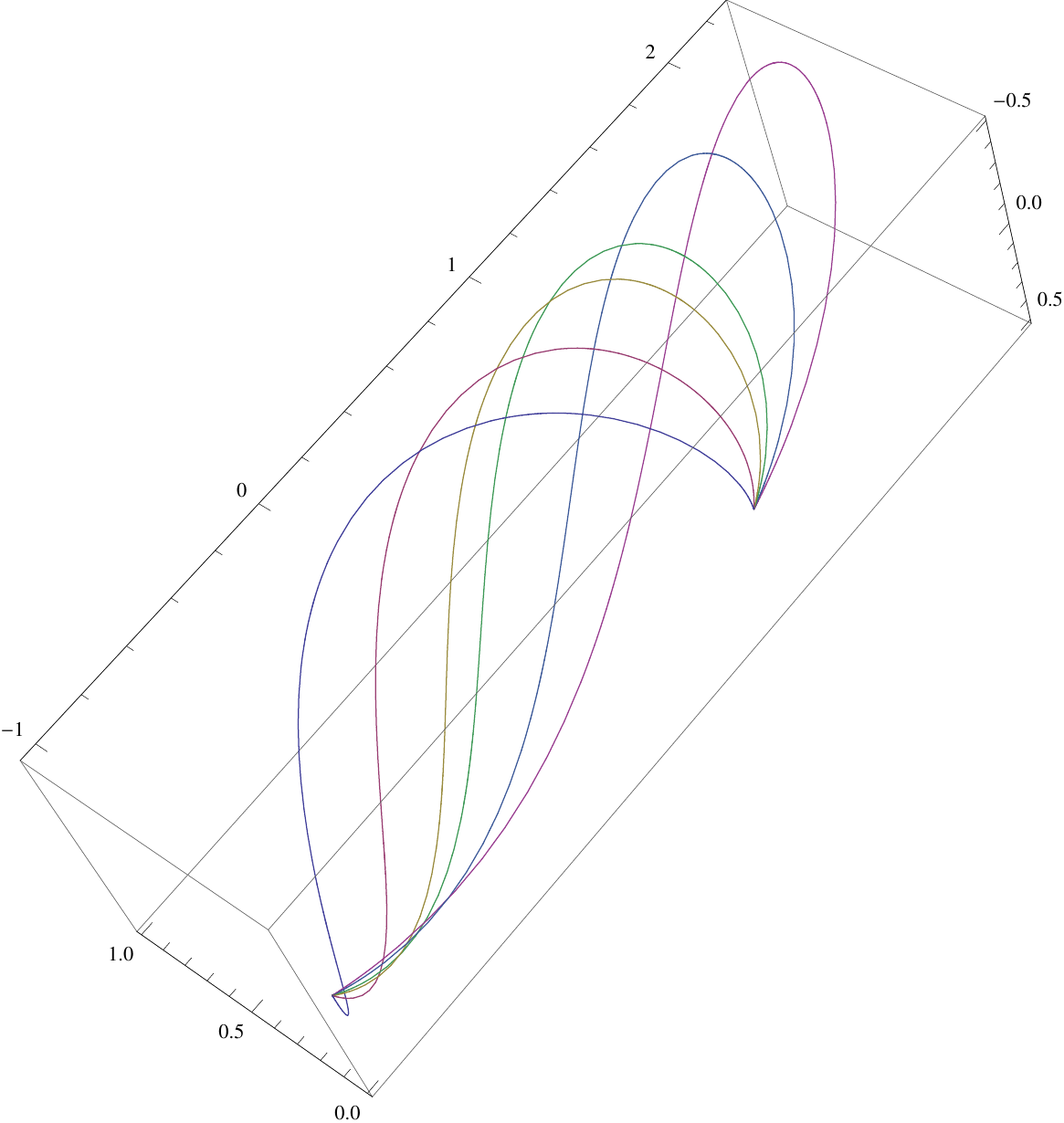}
\caption{Null geodesics when $a=M=K=1$ and $b=-1$ with $t$ along the longest side of the box and different values of $c$.}
\end{figure}

Case 2 : Null geodesics, $a=-\alpha^2<0$, $b=\beta^2>0$. We can solve the integral
(\ref{jgvghjc}) with $s=0$ to get (setting $\phi_0=0$)
\begin{eqnarray}
r &=& \frac{\mathrm{e}^{\phi/\alpha}}{2\,K^2} - \frac{M^2\,\beta^2\,\mathrm{e}^{-\phi/\alpha}}{2},\quad 
t/r = \frac{\alpha  K \left(e^{\frac{2 \phi }{\alpha }}+\beta ^2 K^2 M^2\right)}{\beta ^4 K^2 M^3-\beta ^2 M
   e^{\frac{2 \phi }{\alpha }}}+c\ .
\end{eqnarray}
As $\phi$ varies, we get a spiral, starting at $r=0$ with $\phi=\alpha\,\log_e(M\,K\,\beta)$ and with $r\to\infty$ as $\phi\to\infty$. 

\subsection{Inversion and null geodesics}
We recall that one of the two metrics singled out
by the energy-momentum tensor being that of a perfect fluid
(see Case \ref{vhjkakjg}) was $1=a=\alpha^2>0$, $b=-\beta^2<0$.
The null geodesics are given by putting $\alpha=1$ in 
(\ref{kjhgcvxfggs}) to get (setting $\phi_0=0$)
\begin{eqnarray} \label{kjhgcvxfsdggs}
r &=& \frac{M\,\beta}{K}\,\sin(\phi)\ ,\quad t/r = -\,\frac{K\,\cot(\phi)}{\beta^2\,M}+c\ .
\end{eqnarray}
We will now perform an inversion of the geometry  to a new radial coordinate $\hat r =1/r$, and a new time coordinate $\hat t=t/r$. Then in terms of the new $\hat x,\hat y,\hat z$ coordinates (using the new radius $\hat r$) we get 
\begin{eqnarray*}
(\hat t,\hat x,\hat y,\hat z)=(-\,\frac{K\,\cot(\phi)}{\beta^2\,M}+c,\frac{K\,\cot(\phi)}{M\,\beta},\frac{K}{M\,\beta},0)
\end{eqnarray*}
In other words, we have a straight line in the $\hat x,\hat y$ plane being traversed at constant speed $\beta$ with respect to $\hat t$. If we have nonzero $\phi_0$, the only effect is to rotate this picture, so the general description remains true. 

Now we use the new coordinates $\hat t,\hat r$, together with the usual angular coordinates (totalling $\hat t,\hat r,\theta,\phi$ in that order) to give a change of coordinates,
in which the metric becomes
\begin{eqnarray}\label{newmetricclass}
\hat g_{ij}\ =\ \frac{1}{\hat r^4}\ \left(\begin{array}{cccc}b & 0 & 0 & 0 \\0 & a & 0 & 0 \\0 & 0 & \hat r^2 & 0 \\0 & 0 & 0 & \hat r^2\,\sin^2(\theta)\end{array}\right)\ .
\end{eqnarray}

\subsection{Geodesics in the 2D case}

Here we look in more detail at the radial-time sector of the geodesic motion where the angular momentum $K=0$ and (say) $\phi=0,\theta={\pi\over 2}$ identically. These formulae also apply to the $1+1$ case with the difference that $r$ is allowed to be negative as the Cartesian space coordinate. We take $a=1$ and $b<0$ corresponding to Minkowski signature. From the Ricci tensor we know that we have a  singularity on the line $r=0$. 

In this case the null geodesic equation with $s=0$ in (\ref{trgeodesic}) gives
\[ t=r c- {M\over \sqrt{-b M^2}}\]
which depends on $M$ only through its sign. We take the positive square root as a choice in the affine parameter. The geodesics are straight lines of slope $c$ all passing through points 
\[ P_\pm=(0,\pm {1\over \sqrt{-b}})\]
 on the singularity according to the sign of $M$, as shown for $b=-1$ in Figure~2(a). At every point $P$ other than $P_\pm$ there is precisely one null geodesic with $M=1$ and  one with $M=-1$ passing through that point, as expected. We think of a geodesic as emerging from  from $P_-$ in the past light cone of $P$ and terminating at  $P_+$ in the future light cone.  
\begin{figure}
\[{\rm (a)}\includegraphics[scale=0.5]{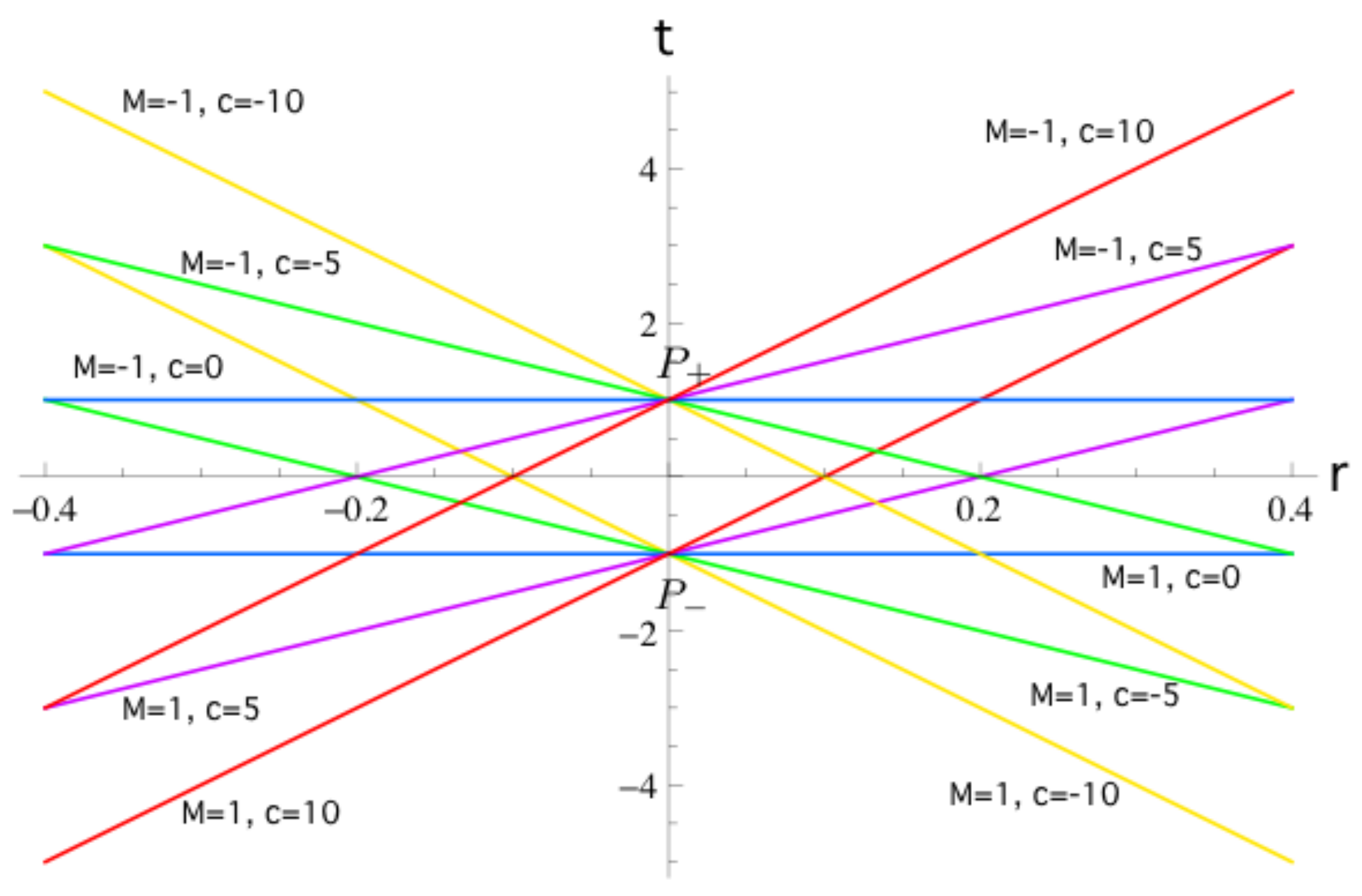}\]
\[{\rm (b)}\includegraphics[scale=0.5]{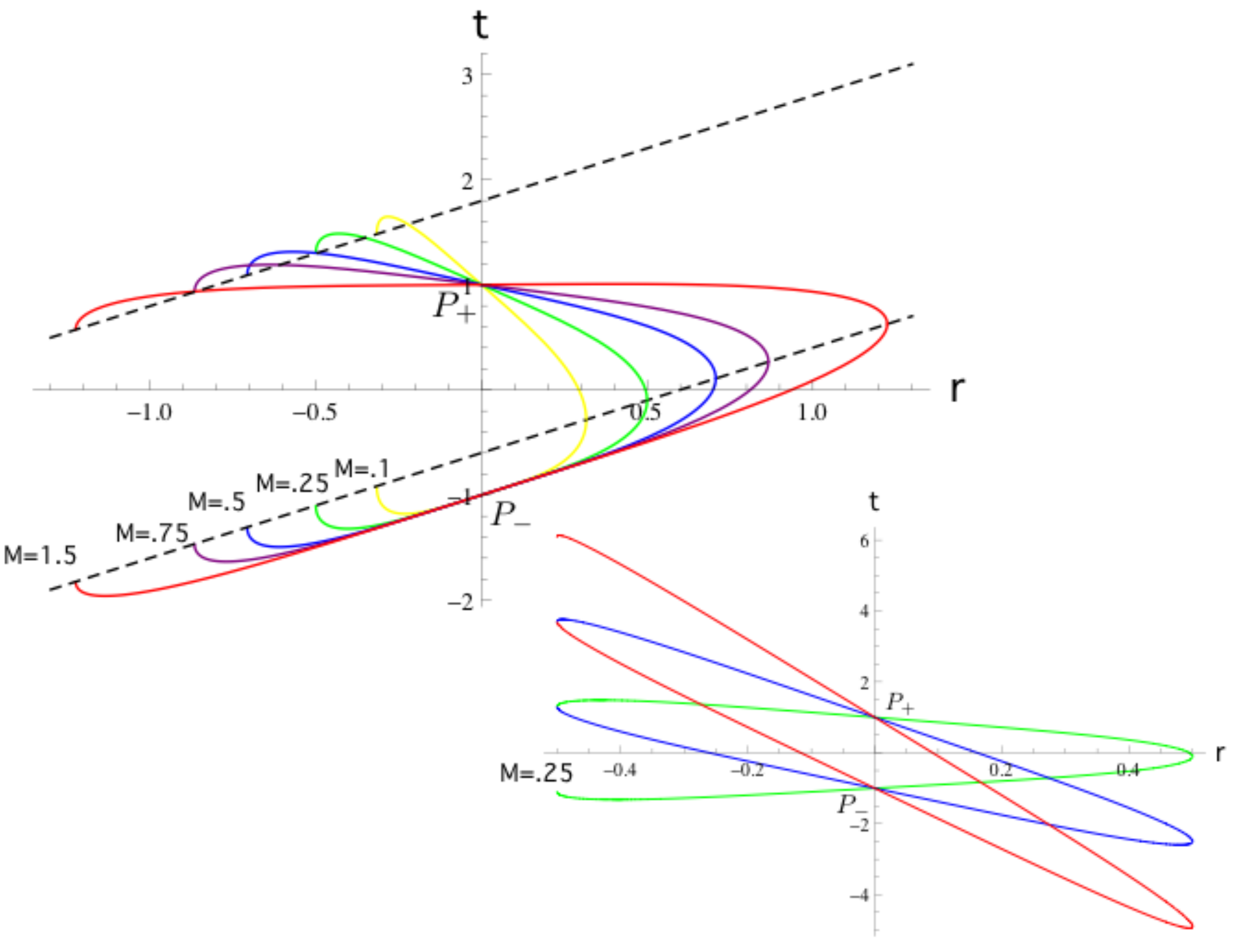}\]
\caption{Classical geodesics in 2D with $a=1$,  $b=-1$. They all pass through one of the two fixed points $P_\pm=(0,\pm 1)$. (a) The null geodesics are shown for $M=\pm 1$ and different slopes $c$.  (b) Timelike geodesics bounce between radius bounds. On bottom right we show 3 complete circuits of a single geodesic with $M=0.25$ and `slope parameter' $c=1$. On bottom left we show the first complete circuit for a range of $M$ all with slope parameter $c=1$. As $M\to \infty$ we obtain two parallel null geodesics of slope $c$.}
\end{figure}

In the case of timelike geodesics with $s=-1$ in (\ref{trgeodesic}) we have
\[ t=r(c\pm M \int {\extd r\over r^2\sqrt{-b {M^2}-r^4}})\]
depending in the branch. We start with a geodesic where we take the positive branch. This is solved as an elliptic function,
\[ t(r)= r c -{M\over D^2}\sqrt{1-({r\over D})^4}-{M \over D^3}  r E({r\over D}),\quad\forall |r|\le D:=(-b M^2)^{1\over 4},\]
\[  E(x)=\int_0^x {u^2\over \sqrt{1-u^4}}\extd u={\rm Elliptic}_{E,-1}(\arcsin(x))-{\rm Elliptic}_{F,-1}(\arcsin(x))\]
in the notation of Mathematica, where $c$ is the constant of integration. We call $c$ the `slope parameter' as it is the slope at the midpoint where the geodesic segment  passes through $r=0$. The value of $M$ is a constant of motion and the geodesics here are parametrized by $M,c$. Also the proper time in this branch according to (\ref{tau}) depends only on $r$. The proper time from $r=-D$, say, is
\[ \tau(r)= D\gamma+\int^r_0 {s^2\over \sqrt{D^4-s^4}}\extd s=D\gamma+ D E({r\over D})\]
independent of the slope $c$ and of the sign of $M$. Here $D\gamma$ is the proper time from  the left boundary at $r=-D$ to $r=0$ while
\[ \gamma=E(1)=\sqrt{\pi}{\Gamma({3\over 4})\over\Gamma({1\over 4})}\approx 0.59907.\]
In principle, we can now write $t,r$ as functions of $\tau$ via the inverse function to the elliptic function $E$ (we have not found a closed formula for this). The proper time from $r=0$ to the right boundary at $r=D$ is again $D\gamma$. The above geodesic segment starts at $\tau=-D\gamma$ at $(-D,-Dc-{M\over D^2}\gamma)$ and ends at $\tau=D\gamma$ at $(D,Dc-{M\over D^2}\gamma)$. At half way it passes through $(0,- {M\over D^2})=P_\mp$ depending on the sign of $M$. 

Motion is necessarily bounded in the region $|r|\le D$ (for the solutions to exist in view of (\ref{trgeodesic})) and as the above geodesic segment approaches the $r=D$ boundary the value of 
\[ \ddot r=2 b {M^2\over r^5}\]
from (\ref{ddot r}) is finite and retains its negative sign while $\dot r\to 0$ in the finite proper time. It follows that motion bounces off the boundary and continues with the reversed sign of the branch of the square root so that the radius is now decreasing with proper time. This starts a new geodesic segment which passes through $P_+$, so in particular we have a timelike geodesic from $P_-$ to $P_+$. If we allow negative $r$ then the geodesic continues and ends at $r=-D$ where the geodesic bounces of the boundary and continues back with the positive branch, and so on. To compute these further segments we solve with the reversed sign of the square root, matching the start point of each segment with the end point of the previous. As $\dot r$ and $\dot t$ depend only in $r$ and the branch, these also match up. The solution above thus becomes the segment $t_0$ of a sequence:
\[ t_n(r)=r c-2n{M\gamma\over D^{3}} r- (-1)^n{M\over D^2}\left(\sqrt{1-({r\over D})^4}+{ r\over D} E({r\over D})\right),\quad\forall |r|\le D, \quad n\in\Z,\]
for the segment where
\[  n 2 D\gamma\le \tau\le (n+1)2D\gamma.\]
The entire solution is plotted at right in Figure~2(b) for three `full cycles' from the $r=-D$ boundary and back. Notice the `precession' whereby the slope takes a negative step with each full cycle. 

Meanwhile, at left in Figure~2(b) we plot just one `full cycle' (i.e. the initial segment $t_0$ and the next  segment $t_1$) but for increasing values of positive $M$. The end points after a full cycle lie on the upper dashed line while the start  points are on the lower dashed line according to the analysis above. As $M\to \infty$ these two segments limit for any bounded region of radius to a pair of null geodesics of the same slope $c$. This gives a picture of the parallel slopes in Figure~2(a); we think of the lower one passing through $P_-$ as a light ray heading out to the boundary at infinity where it bounces back and becomes the upper one passing through $P_+$. This picture is also consistent in 3+1 with an appropriate limit as $K\to 0$ of the null geodesics in Figure~1, as these connect $P_\pm$. Our interpretation is for $M>0$. When $M<0$ our above solution for a geodesic is the $t$-reverse of the same solution at positive $M$ but reversed value of $c$. 

The above physical picture of the geodesics in the case $a=1,b<0$ represents a gravitational source at $r=0$ which is so strong that all outgoing geodesics eventually come back. This is perhaps clearer using the $\hat t=t/r$ variable from Section~3.3. Then
\[ g= b r^4 \extd \hat t^2+a\extd r^2 +  r^2\,(\extd\theta^2+\sin^2\theta\,\extd\phi^2)\] 
is spatially the flat metric but with a `Newtonian potential' term $\beta^{-1}\extd \tilde t^2$ cf.\cite{Ma:newt}, where in our case $\beta={1 \over  b r^4}$ . The special points $P_\pm$ now go to $\pm\infty$ in the $\hat t$-axis.   As we have seen in Section~3.2, the Einstein tensor corresponds when $a=1$ and $n=4$ to some kind of perfect fluid, albeit an unphysical one of positive pressure zero density, while when $n=2$ we do not need any matter.  

\begin{figure}
\[\includegraphics[scale=0.4]{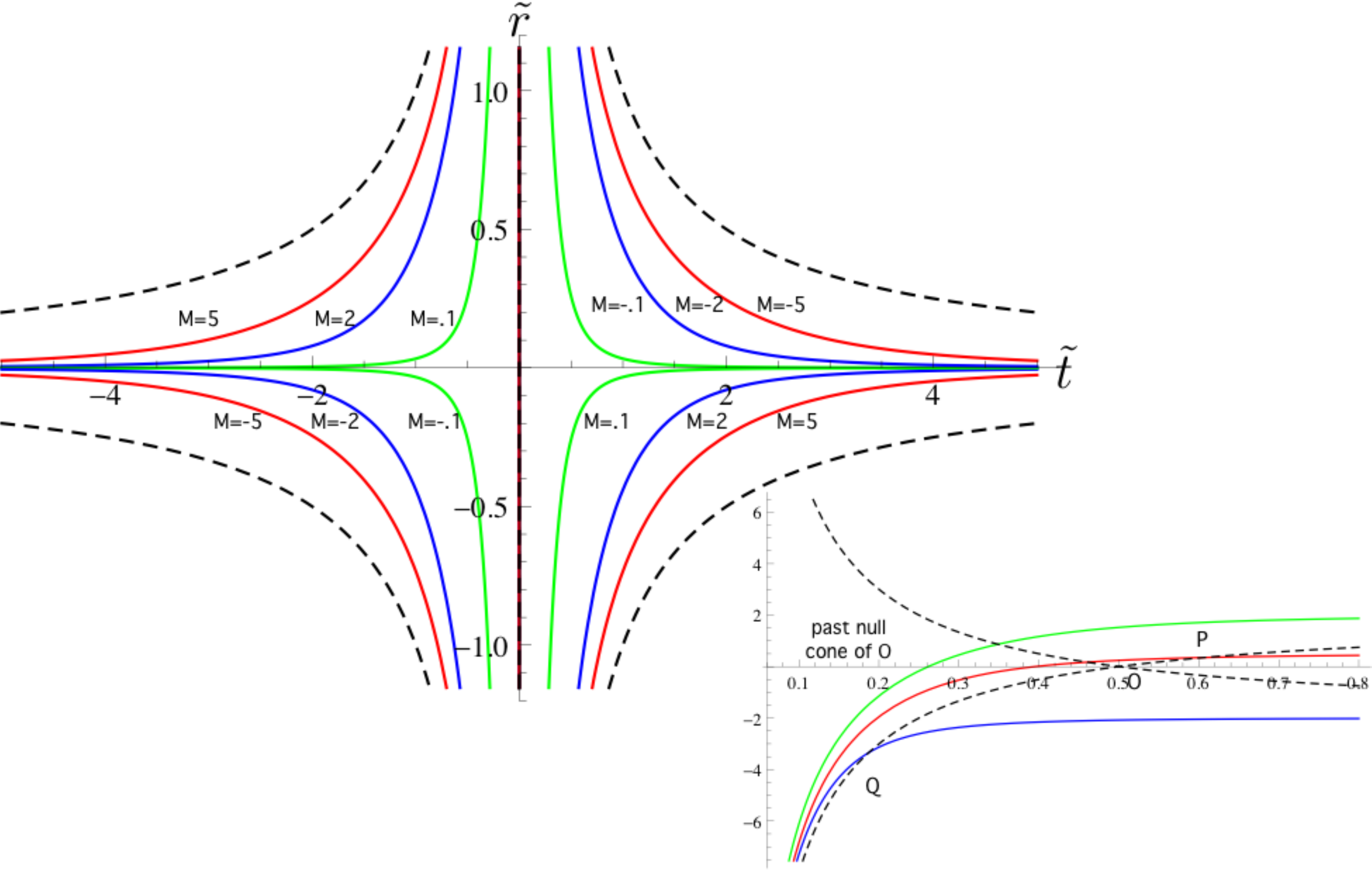}\]
\caption{Classical timelike geodesics in the 2D FRW-like coordinates with $a=-1$,  $b=1$ (they are spacelike for the previous intepretation). Direction of travel with respect to proper time is to the right.  The constant of integration $c$ simply shifts the curves vertically and in the main picture has been chosen so that all geodesics asymptote to $0$ as $\tilde t\to \pm\infty$. Dashed lines in the smaller picture show the light cone at a point $O$. Strictly timelike geodesics never cross or cross the corresponding null-geodesic through $O$ at some point $Q$ or $P$, but in all cases have been in the past light cone as shown. }
\end{figure}

When $n=2$ one also has an FRW-like interpretation when $a<0$ and $b>0$, using $\tilde t=r$ and $\tilde r=\hat t$ as new variables. Then
\[ g=a\extd \tilde t^2+R(\tilde t)^2 \extd \tilde r^2,\quad R(\tilde t)=\sqrt b\, \tilde t^2.\]
and now the Ricci singularity is on the $\tilde r$-axis at $\tilde t=0$. Timelike geodesics are the  spacelike ones in the previous analysis and are shown in Figure~3. Normalising so that $a=-1$, these geodesics are solutions of 
\[ \ddot{\tilde t}=-2 {\tilde t}^3 b {\dot{\tilde r}}^2,\quad \ddot {\tilde r}=-4{\dot{\tilde r}\, \dot{\tilde t}\over \tilde t},\quad {\dot{\tilde t}}=\sqrt{1+ b{ \tilde t}^4\,  {\dot{\tilde r}}^2}\]
from which we see that $M:={\tilde t}^4\dot{\tilde r}$ is a constant of motion. We fix here the positive square root (say) so that $\tilde t$ is always increasing with proper time. Note also that $\tilde r$ is increasing when $M>0$ and decreasing when $M<0$ with the result that as proper time increases we always move to the right along the timelike geodesics in the figure. The exact solution up to a constant of integration is provided this time by hypergeometric functions
\[ \tilde{r} =c - {M\over 3 {\tilde t}^3} ({}_2F_1)({1\over 2}, {3\over 4}, {7\over 4}, -{b M^2\over{\tilde t}^4})\]
in the notation of Mathematica, where $c$ is the value of $\tilde r$ at $\tilde t=\pm\infty$. For positive $\tilde t$, geodesics  start arbitrarily close to $\tilde t=0$ and $\tilde r=\mp \infty$ (according to the sign of $M$) and asymptote to a constant value $c$ in a finite proper time from any finite point (but the proper time from $\tilde r=\mp\infty$ is infinite).  The limit of infinite $|M|$ gives null geodesics (dashed in the figure, they are  hyperbolae $\tilde r= c \mp {1\over \sqrt{b}\tilde t}$). The forward light cone of any point $O=({\tilde t}_O,{\tilde r}_O)$ where ${\tilde t}_O>0$ spans at $\tilde t=\infty$ only the finite band in $\tilde r$ between ${\tilde r}_O\pm {1\over \sqrt{b}{\tilde t}_O}$. Meanwhile, as shown in the lower part of Figure~3, strictly timelike geodesics either stay above the corresponding null geodesic or cross it before or after  (or perhaps at)  $O$, and in all cases have already been in the backward light cone. Since they can be pitched to asymptote to any value of $\tilde r$ and hence any distance from $O$, the particle horizon at $O$ is infinite (there is no horizon in the sense of a limited range above and below $O$ through which pass all geodesics from $\tilde t=0$ that pass through the backward light cone). 

One can vertically compress the picture by using $\tilde t\tilde r$ in place of $\tilde r$, in which case timelike geodesics for $\tilde t\ge 0$ begin at finite points $P_\pm=(0, \pm{ 1\over\sqrt{b}})$ and null geodesics are again straight lines with slope $c$. Again because we are in $n=2$ we do not need any matter source.

\section{A noncommutative change of basis}

Although we are not going to explore any serious noncommutative geometry for $n>2$, we give here a small application motivated by our above analysis of geodesics on our curved classical metric. Namely, we show that the  change of coordinates suggested in Section~3.3 works and gives a nice answer in the quantum algebra.  Now the $t,x_i$ for  $i=1,\cdots,n-1$ are generators of the bicrossproduct model quantum spacetime algebra (\ref{mink}) and we set 
\begin{eqnarray}
\hat x_i\,=\, r^{-2}\,x_i\ ,\quad \hat t\,=\,\tfrac12 (r^{-1}\,t+t\,r^{-1})\ .
\end{eqnarray}
Then $\hat r^2=\sum_i \hat x_i^2=r^{-2}$, and
\begin{eqnarray*}
[\hat x_i,t] &=& r^{-2}\,[x_i,t] + [r^{-2},t]\,x_i \, =\, \lambda\,r^{-2}\,x_i-2\,\lambda\,r^{-2}\,x_i \,=\, -\lambda\,r^{-2}\,x_i\ ,\cr
2\,[\hat x_i,\hat t] &=& r^{-1}\,[\hat x_i,t]+[\hat x_i,t]\,r^{-1} \,=\, -2\,\lambda( r^{-3}\,x_i  )\ ,\cr
2\,[\hat r,\hat t] &=& r^{-1}\,[r^{-1},t]+[r^{-1},t]\,r^{-1}\ =\  -2\,\lambda\, r^{-2}\ .
\end{eqnarray*}
Thus we have {\em quadratic} commutation relations 
\begin{eqnarray}
[\hat x_i,\hat x_j]\,=\,0\ ,\quad [\hat x_i,\hat t] \,=\, -\lambda\,\hat r\,\hat x_i
\ ,\quad [\hat r,\hat t] \,=\, -\lambda\, \hat r^2  
\end{eqnarray}
for the algebra and the additional quadratic relation  $\hat r^2=\sum_i\hat x_i^2$. 

For the differential calculus, the relation $[\extd x_i,x_j]=0$ gives $[\extd \hat x_i,\hat x_j]=0$. Next
\begin{eqnarray*}
\extd(r^{-1}\,t) &=& r^{-1}\,\extd t-r^{-2}\extd r\,t\ =\ r^{-2}\,(r\,\extd t-t\,\extd r)\,=\, r^{-2}\,v\ ,\cr
\extd(t\,r^{-1}) &=& v^*\, r^{-2}\ .
\end{eqnarray*}
We get for any  $y$ in the algebra
\begin{eqnarray*}
2\,[\extd\hat t,y] \,=\, (v+v^*)\,[r^{-2},y]+[v+v^*,y]\,r^{-2}\ ,
\end{eqnarray*}
and this immediately gives $[\extd\hat t,\hat r]=0$ and $[\extd\hat t,\hat x_i] = \lambda\,(\hat x_i\,\extd\hat r-\hat r\,\extd\hat x_i)$ as well as 
\begin{eqnarray*}
2\,[\extd\hat t,\hat t] \,=\, (v+v^*)\,[r^{-2},\hat t]\,=\, -2\,(v+v^*)\,\lambda\,\hat r^3=-2\,\lambda\,\hat r\,\extd\hat t.
\end{eqnarray*}
Here
\[v+v^* = 2\,\hat r^{-2}\,\extd\hat t\ ,\quad v^*-v = \lambda\,\hat r^{-2}\,\extd\hat r\ \]
\[ v^* = \hat r^{-2}\,\extd\hat t+\tfrac\lambda2\,\hat r^{-2}\,\extd\hat r\ ,\quad 
v = \hat r^{-2}\,\extd\hat t-\tfrac\lambda2\,\hat r^{-2}\,\extd\hat r\ .\]
Next
\begin{eqnarray*}
[\extd\hat x_i,t] &=& [r^{-2}\,\extd x_i-2\,r^{-3}\,\extd r\,x_i,t] \ ,
\end{eqnarray*}
which we compute further to complete the following full set of relations for the differential calculus in these generators:
\begin{eqnarray} [\extd \hat x_i,\hat x_j]&=&0,\quad [\extd\hat t,\hat r]=0,\quad [\extd\hat t,\hat x_i] = \lambda\,(\hat x_i\,\extd\hat r-\hat r\,\extd\hat x_i)\end{eqnarray}
\begin{eqnarray}
 [\extd\hat t,\hat t]&=&-2\,\lambda\,\hat r\,\extd\hat t,\quad  [\extd\hat x_i,\hat t]=-2\,\lambda\,\hat r\,\extd\hat x_i\ .\end{eqnarray}
One can also define $\hat\omega_i=\extd \hat x_i - (\hat x_i/\hat r)\extd\hat r=r^{-2}\omega_i$ after a short computation.

For the quantum metric in these new coordinates, we compute (summing over repeated indices)
\begin{eqnarray*}
\omega_i\tens\omega_i &=& \hat r^{-4}\,(\extd \hat x_i\tens \extd \hat x_i- \extd\hat r\tens \extd\hat r)=\hat r^{-4}\hat \omega_i\tens\hat \omega_i\ ,\cr
\extd r\tens\extd r &=& \hat r^{-4}\,\extd\hat r\tens \extd\hat r\ ,
\end{eqnarray*}
\begin{eqnarray*}
v^*\tens v &=& \hat r^{-4}\,\big( \extd\hat t\tens \extd\hat t +\lambda\,(\extd\hat r\tens \extd\hat  t-\extd\hat  t\tens\extd\hat  r)/2-\lambda^2\, \extd\hat r\tens \extd\hat r/4
\big)\ ,\cr
\extd r\tens v+v^*\tens\extd r &=& -\, \hat r^{-4}\,\big( \extd\hat r\tens \extd\hat  t + \extd\hat  t \tens \extd\hat  r\big)\ ,\cr
\extd r\tens v-v^*\tens\extd r &=& \hat r^{-4}\,\big( \extd\hat t\tens \extd\hat  r - \extd\hat  r\tens \extd\hat  t +\lambda\, \extd\hat r\tens \extd\hat  r
\big)\ .
\end{eqnarray*}
Then the quantum metric in Proposition~\ref{dhjsvak}  becomes
\begin{eqnarray*} g&=&\hat r^{-4}\left(\hat\omega_i\tens\hat\omega_i+(a+{3 b\lambda^2\over 4}) \extd\hat r\tens \extd\hat r +b\extd\hat t\tens \extd\hat t +{b\lambda\over 2}( \extd\hat t\tens \extd\hat  r - \extd\hat  r\tens \extd\hat  t)\right)\end{eqnarray*}
showing a form similar to the classical case (\ref{newmetricclass})  in these variables, with quantum corrections. 

\section{Noncommutative geometry of the 2D model at first order}

Here we completely solve the noncommutative Riemannian geometry of the 2D bicrossproduct model to order $\lambda$ in the deformation parameter. We will write the spatial variable as $r$ in order to make contact with the general case, i.e. thinking also of the model below as a limit of the full metric in say 4D  but with angular modes suppressed. 

The formalism of noncommutative Riemannian geometry on an algebra $A$ that we will use is the constructive one from our paper \cite{BegMa4} and used recently in \cite{Ma:alm}. This is based on bimodule connections\cite{Mou,DV1,DV2} which in the case of a linear connection on the bimodule $\Omega^1$ of 1-forms amounts to a linear map $\nabla:\Omega^1\to \Omega^1\tens_A \Omega^1$  obeying
\[ \nabla(a\omega)=\extd a\tens_A \omega+a\nabla \omega,\quad \nabla(\omega a)=(\nabla \omega)a+\sigma(\omega\tens_A \extd a),\quad\forall a\in A,\ \omega\in\Omega^1\]
for some bimodule map $\sigma:\Omega^1\tens_A \Omega^1\to \Omega^1\tens_A\Omega^1$ called the `generalised braiding' (in some cases it obeys the braid relations). The notion of connection here is similar to that of a covariant derivative $\nabla_X$ except that the first tensor factor of the output of $\nabla$ is a copy of $\Omega^1$ waiting to be evaluated on a vector field.  The map $\sigma$ is needed to flip factors in order for this interpretation to make sense, and classically it is a flip. In particular, we formulate a metric as a nondegenerate element $g\in\Omega^1\tens_A\Omega^1$ and now the notion of metric compatibility makes sense as
\begin{equation}\label{metriccompat} \nabla g\equiv (\nabla\tens \id)g+( \sigma\tens\id)(\id\tens\nabla)g=0.\end{equation}
The notion of torsion free also makes sense, as $\wedge \nabla=\extd$ provided $\Omega^2$ is defined. Hence there is a notion of `quantum Levi-Civita connection'. 

In our case the quantum metric  from Proposition~\ref{dhjsvak} up to an overall normalisation now has the reduced form
\begin{equation}\label{metric2d} g=\extd r\tens\extd r+ b\, (v^*\tens v + \lambda (\extd r\tens v-v^*\tens\extd r)) \end{equation}
for a single real parameter $b$ (we have set $a=1$). 

From Section~\ref{bnjlilygov} we have the classical Levi-Civita covariant derivative, which becomes the $O(\lambda^0)$ part of the noncommutative covariant derivative
\[ \nabla_0(\extd r) =\frac{2\,b}{r}\,v\tens v, \quad
\nabla_0(v) =  -\,\frac{2}{r}\,v\tens \extd r\ \]

We wish to extend this calculation to $O(\lambda)$ in the noncommutative case.
We take $\nabla=\nabla_0+\lambda\,\nabla_1+O(\lambda^2)$. 
 The first task is to calculate $\sigma$ assuming it exists, which we do from the formula
\begin{eqnarray}
\sigma(\omega\tens\extd a) \,=\, \extd a\tens\omega+[a,\nabla(\omega)]+\nabla([\omega,a])\ .
\end{eqnarray}
Notice that to $O(\lambda)$ it is enough to calculate this using $\nabla_0$, which gives the result:
\begin{eqnarray*}
\sigma(\omega\tens\extd r) & = & \extd r\tens\omega\ , 
\end{eqnarray*}
for $\omega$ any of $\extd r,v$. Also
\begin{eqnarray*}
\sigma(\extd r\tens\extd t) & = & \extd t\tens\extd r   +[t,\nabla_0(\extd r)] = \extd t\tens\extd r +  \frac{2\,b\,\lambda}{r}\,v\tens v   \ ,  \cr
\sigma(v\tens\extd t) & = & \extd t\tens v  +[t,\nabla_0(v)] = \extd t\tens v -\frac{2\,\lambda}{r}\,v\tens \extd r   \ .  \cr
\end{eqnarray*}
Now we have
\begin{eqnarray*}
\sigma(\omega\tens v^*) &=& \sigma(\omega\tens \extd t).r-\sigma(\omega\tens \extd r).t \ ,\cr
\sigma(v\tens v^*) &=& \extd t\tens v.r-\frac{2\,\lambda}{r}\,v\tens \extd r.r -\extd r\tens v.t \cr
&=& v^*\tens v-2\,\lambda\,v\tens \extd r\ ,\cr
\sigma(\extd r\tens v^*) &=& v^*\tens\extd r +{2\,b\,\lambda}\,v\tens v\ .\cr
\end{eqnarray*}
as the braiding to $O(\lambda)$. Summarising in terms of $v$, we have to order $\lambda$,
\begin{eqnarray}\label{braiding2d}
\sigma(v\tens v) &=&  v\tens v-2\,\lambda\,v\tens \extd r\ ,\quad \sigma(\extd r\tens v) = v\tens\extd r +{2\,b\,\lambda}\,v\tens v\ ,
\cr\sigma(v\tens \extd r) &=&  \extd r\tens v,\quad \sigma(\extd r\tens \extd r) = \extd r\tens\extd r \ .
\end{eqnarray}

Next we will find the connection effectively using a `Koszul formula' in \cite{BegMa4}. This method makes essential use of the $*$-operation so we need to explain this first. Recall that in noncommutative geometry we do not work with the analogue of real-valued functions on a manifold but complex valued ones, generalised now to a $*$-algebra. In the commutative case one may recover a real subalgebra by looking at hermitian elements where $a^*=a$ but in general one may not have this luxury. The same applies to the differential forms where we extend $*$ to an operation on the exterior algebra. Now that we are working over $\C$ a metric $g$ is in principle complexified but we can impose a `hermitian' condition as explained in Section~2 as a form of reality constraint.  We need to explain similarly the correct $*$-preserving or `reality' property of a bimodule connection, a problem which was solved in general in \cite{BegMa4}. 

We will explain this only in the case of $\Omega^1$ needed here, but the general case similar. The first step is to define a conjugate bimodule $(\overline{\Omega^1},\cdot) $ which is the same abelian group under addition as $\Omega^1$ but taken with a conjugate action $a.\bar\omega=\overline{\omega a^*}$ and $\overline\omega\cdot a=\overline{a^*\omega}$ for all $a\in A$, $\omega\in \Omega^1$ and $\bar\omega$ the same element viewed in $\overline{\Omega^1}$. The conjugate includes the action of scalars in $A$. We view $*$ itself more properly as a bimodule map $\star:\Omega^1\to \overline{\Omega^1}$. Another ingredient is a map $\Upsilon: \overline{\Omega^1}\tens\overline{\Omega^1}\to \overline{\Omega^1\tens\Omega^1}$ which in our case is just the flip map but with elements viewed appropriately. Using conjugate modules one may formulate a notion of a connection $\nabla$ being star-preserving\cite{BegMa4}, which in our case  for $\xi^*\in\Omega^1$ amounts to 
\begin{eqnarray*}
(\id\tens\star)\nabla\,\star^{-1}(\overline{\xi^*}) &=& (\star^{-1}\tens\id) \Upsilon\,\overline{\sigma^{-1}
\nabla(\xi^*)}\ ,\cr
(\id\tens\star)\sigma(\star^{-1}\tens\id)&=&  (\star^{-1}\tens\id) \Upsilon\, \overline{\sigma^{-1}}\, \Upsilon^{-1}
(\id\tens\star)\ ,
\end{eqnarray*}
and rearrangement of this gives
\begin{eqnarray} \label{bvdhskalvbk}
\nabla(\xi) &=& \sigma(\star^{-1}\tens\star^{-1}) \Upsilon\,\overline{
\nabla(\xi^*)}\ ,\cr
(\star\tens\star)\sigma(\star^{-1}\tens\star^{-1})&=&  \Upsilon\, \overline{\sigma^{-1}}\, \Upsilon^{-1}\ .
\end{eqnarray}
or in concrete terms 
\begin{eqnarray*}
\nabla(\xi) &=& \sigma(\zeta^*\tens\eta^*),\quad \forall \eta\tens\zeta=\nabla(\xi^*).\end{eqnarray*}

To analyse this we set $\nabla=\nabla_0+\lambda\,\nabla_1$, and use the fact that we know $\sigma$ to $O(\lambda)$ already. We set $\eta_0\tens\zeta_0=\nabla(\xi^*)$ and $\eta_1\tens\zeta_1=\nabla_1(\xi^*)$, and
\begin{eqnarray*}
\nabla_0(\xi) +\lambda\, \nabla_1(\xi) &=& \sigma(\zeta_0^*\tens\eta_0^*)-\lambda\,\sigma(\zeta_1^*\tens\eta_1^*)\ ,
\end{eqnarray*}
and as $\sigma$ is just transpose to $O(\lambda^0)$ we get to $O(\lambda^1)$
\begin{eqnarray}   \label{cxbvhaku}
\nabla_0(\xi) +\lambda\, \nabla_1(\xi) &=& \sigma(\zeta_0^*\tens\eta_0^*)-\lambda\,\eta_1^*\tens\zeta_1^*\ .
\end{eqnarray}
In our cases, we have $\xi^*=\xi$ to $O(\lambda^0)$, so to $O(\lambda)$, $\lambda\, \nabla_1(\xi)=\lambda\,\eta_1\tens\zeta_1$, so (\ref{cxbvhaku}) becomes
\begin{eqnarray}   \label{cxbvhakub}
\lambda\,(\eta_1\tens\zeta_1+\eta_1^*\tens\zeta_1^*) &=& \sigma(\zeta_0^*\tens\eta_0^*)-\nabla_0(\xi) \ .
\end{eqnarray}

\medskip
Case 1: $\xi=\extd r$, $\xi^*=\xi$, and then 
\begin{eqnarray*}
\eta_0\tens\zeta_0 &=&  \frac{2\,b}{r}\,v\tens v \cr
\zeta_0^*\tens\eta_0^* &=&  \frac{2\,b}{r}\,v^*\tens v^* \cr
 &=& \frac{2\,b}{r}\,v\tens v^*-  \frac{2\,b\,\lambda}{r}\,\extd r\tens v^* \cr
 \sigma(\zeta_0^*\tens\eta_0^*) &=&   \frac{2\,b}{r}\,v^*\tens v
 - \frac{6\,\lambda\,b}{r}\,v\tens\extd r
\end{eqnarray*}
and substituting this in (\ref{cxbvhakub}) gives to $O(\lambda^0)$, where $\eta_1\tens\zeta_1=\nabla_1(\extd r)$
\begin{eqnarray*}
\eta_1\tens\zeta_1+\eta_1^*\tens\zeta_1^*\,=\,-\frac{6\,b}{r}\,v\tens \extd r - \frac{2\,b}{r}\,\extd r\tens v
\ .
\end{eqnarray*}
Using the notation that $\tau\tens\kappa$ is an $O(\lambda^0)$ Hermitian tensor product,
\begin{eqnarray} \label{khtdcdik2}
\nabla(\extd r)\,= \frac{2\,b}{r}\,v\tens v 
-\frac{3\,b\,\lambda}{r}\,v\tens \extd r - \frac{\lambda\,b}{r}\,\extd r\tens v
 + \mathrm{i}\,\lambda \, \tau_r\tens\kappa_r\ .
\end{eqnarray}

\medskip
Case 2: $\xi=v$, $\xi^*=v-\lambda\,\extd r$, and then 
\begin{eqnarray*}
\eta_0\tens\zeta_0 &=&  -\,\frac{2}{r}\,v\tens \extd r 
-  \frac{2\,b\,\lambda}{r}\,v\tens v \ ,\cr
\zeta_0^*\tens\eta_0^* &=& -\, \frac{2}{r}\,\extd r\tens v^*  +  \frac{2\,b\,\lambda}{r}\,v^*\tens v^*  \ ,\cr
 \sigma(\zeta_0^*\tens\eta_0^*) &=& -\,\frac2r\,\Big(v^*\tens\extd r +{2\,b\,\lambda}\,v\tens v
 \Big)  
+  \frac{2\,b\,\lambda}{r}\,v^*\tens v^* \cr
 &=& -\,\frac2r\,v^*\tens\extd r  -  \frac{2\,b\,\lambda}{r}\,v^*\tens v^* \ .
\end{eqnarray*}
Next 
\begin{eqnarray*}
\nabla(v) &=& \nabla(v^*) +\lambda\,\nabla(\extd r) \cr
&=& \eta_0\tens\zeta_0+\lambda\,\eta_1\tens\zeta_1
+  \frac{2\,b\,\lambda}{r}\,v\tens v
\end{eqnarray*}
and substituting this in (\ref{cxbvhakub}) gives to $O(\lambda^0)$, where $\eta_1\tens\zeta_1=\nabla_1(v^*)$ 
\begin{eqnarray*}
\eta_1\tens\zeta_1+\eta_1^*\tens\zeta_1^*\,=\, \frac{2}{r}\,\extd r\tens\extd r 
-\frac{2\,b}{r}\,v\tens v \ .
\end{eqnarray*}
Now we get
\begin{eqnarray} \label{khtdcdik3}
\quad\nabla(v^*) &=&   -\,\frac{2}{r}\,v\tens \extd r -  \frac{3\,b\,\lambda}{r}\,v\tens v  
+   \frac{\lambda}{r}\,\extd r\tens\extd r +\mathrm{i}\,\lambda\,\tau_v\tens \kappa_v \ ,\cr
\nabla(v) &=& -\,\frac{2}{r}\,v\tens \extd r 
-  \frac{b\,\lambda}{r}\,v\tens v  
+   \frac{\lambda}{r}\,\extd r\tens\extd r +\mathrm{i}\,\lambda\,\tau_v\tens \kappa_v \ .
\end{eqnarray}
Here (\ref{khtdcdik2},\ref{khtdcdik3}) is the quantum covariant derivative to $O(\lambda)$ and constructed in such a way as to be $\star$-preserving to this order.

Next, we use $v\wedge v=\lambda\,r\,\extd t\wedge\extd r$ to see that this $O(\lambda)$ covariant derivative 
is torsion free to $O(\lambda)$, i.e.\ that $\wedge\nabla=\extd$, as long as $\tau_v\wedge \kappa_v=0$ and $\tau_r\wedge \kappa_r=0$. It should have been noted that if the antihermitian $O(\lambda)$ part calculated in (\ref{khtdcdik2},\ref{khtdcdik3}) had come out differently, there would have been no way to correct this to give zero torsion by using the $\tau\wedge\kappa$ terms, as they are all Hermitian. 

Finally, we look at metric compatibility.  If $\nabla$ is $\star$-preserving one can show that metric compatibility is equivalent to Hermitian-metric compatibility of the associated sesquilinear quantum metric
\begin{equation}\label{sesmetric2d} (\star\tens{\id})g=\overline{\extd r}\tens\extd r+ b\, (\overline{v}\tens v + \lambda (\overline{\extd r}\tens v-\overline{v}\tens\extd r)) \end{equation}
to which we apply the covariant derivative as $\nabla_{\overline{\Omega^1}}\tens\id+\id\tens\nabla_{\Omega^1}$. The `hermitian' or reality property of $g$ used in Section~2 also appears more simply in terms of the sequilinear quantum metric as invariance under flip, where we identify the barred and unbarred spaces and complex-conjugate any coefficients. At least ignoring the optional $\tau\tens\kappa$ terms, we find that 
\begin{eqnarray*}
\big(\nabla_{\overline{\Omega^1}}\tens{\id}+{\id}\tens\nabla_{\Omega^1}\big)\big((\star\tens{\id})g\big) \,=0\end{eqnarray*}
to order $\lambda$, an easier computation as we do not have to deal with the braiding.  It follows that $\nabla g=0$ in the original sense (\ref{metriccompat}) as well to this order, something
that can be also checked directly by a tedious computation.  We summarise the above results:

\begin{proposition}\label{nabla2d} To order $\lambda$,
\begin{eqnarray*} 
\nabla(\extd r)\,&=& \frac{2\,b}{r}\,v\tens v 
-\frac{3\,b\,\lambda}{r}\,v\tens \extd r - \frac{\lambda\,b}{r}\,\extd r\tens v\cr
 \nabla(v) &=& -\,\frac{2}{r}\,v\tens \extd r 
-  \frac{b\,\lambda}{r}\,v\tens v  
+   \frac{\lambda}{r}\,\extd r\tens\extd r 
\end{eqnarray*}
is a bimodule connection on $\Omega^1$ with braiding (\ref{braiding2d}) which is $*$-preserving, torsion free and metric compatible with (\ref{metric2d}) to this order. 
\end{proposition}

Next, the curvature of any left  linear connection in our formalism is given by 
\begin{equation}\label{curvature} R:\Omega^1\to \Omega^2\tens\Omega^1,\quad R=(\extd\tens\id-(\wedge\tens\id)(\id\tens\nabla))\nabla.\end{equation}
We calculate this for our connection using $v\wedge v=\lambda\,v\wedge\extd r$,
\begin{eqnarray}
R(v) &=&  -\,\extd\Big(\frac{2}{r}\,v\Big)\tens \extd r 
-  \extd\Big(\frac{b\,\lambda}{r}\,v\Big)\tens v  
+  \extd\Big( \frac{\lambda}{r}\,\extd r\Big)\tens\extd r \cr
&&  +\,\frac{2}{r}\,v\wedge\nabla(\extd r )
+  \frac{b\,\lambda}{r}\,v\wedge\nabla( v  )
-   \frac{\lambda}{r}\,\extd r\wedge\nabla(\extd r) \cr
&=&  \,2\,	\frac{v\wedge \extd r}{r^2}\tens \extd r 
+ b\,\lambda\, \frac{v\wedge\extd r}{r^2}\tens v  \cr
&&  +\,\frac{2}{r}\,v\wedge\nabla(\extd r )
+  \frac{b\,\lambda}{r}\,v\wedge\nabla( v  )
-   \frac{\lambda}{r}\,\extd r\wedge\nabla(\extd r) \cr
&=&  \,2\,	\frac{v\wedge \extd r}{r^2}\tens \extd r 
+ b\,\lambda\, \frac{v\wedge \extd r}{r^2}\tens v  \cr
&&  +\,\frac{2}{r}\,v\wedge  \Big(    \frac{2\,b}{r}\,v\tens v 
-\frac{3\,b\,\lambda}{r}\,v\tens \extd r - \frac{\lambda\,b}{r}\,\extd r\tens v  \Big) \cr
&&+\,  \frac{b\,\lambda}{r}\,v\wedge\Big(   -  \frac{2}{r}\,v\tens \extd r  \Big)
-   \frac{\lambda}{r}\,\extd r\wedge\frac{2\,b}{r}\,v\tens v  \cr
&=&  \,2\,	\frac{v\wedge\extd r}{r^2}\tens \extd r 
+ b\,\lambda\, \frac{v\wedge \extd r}{r^2}\tens v  \cr
&&  +\,\frac{2}{r}\,v\wedge  \Big(    \frac{2\,b}{r}\,v\tens v 
 - \frac{\lambda\,b}{r}\,\extd r\tens v  \Big)   -   \frac{2\,b\,\lambda}{r^2}\,\extd r\wedge v\tens v  \cr
 &=&  \,2\,	\frac{v\wedge\extd r}{r^2}\tens \extd r 
+ 3\,b\,\lambda\, \frac{v\wedge \extd r}{r^2}\tens v  \cr
&&  +\,\frac{4\,b}{r^2}\,v\wedge v\tens v 
 -\frac{2\,\lambda\,b}{r^2}\,v\wedge  \extd r\tens v   \cr
  &=&  2\,	\frac{v \wedge \extd r}{r^2}\tens \extd r 
+5\,b\,\lambda\, \frac{v\wedge \extd r}{r^2}\tens v  \ .
\end{eqnarray}
Also we have
\begin{eqnarray}
R(\extd r) &=& \extd\Big(\frac{2\,b}{r}\,v\Big)\tens v 
-\extd\Big(\frac{3\,b\,\lambda}{r}\,v\Big)\tens \extd r - \extd\Big(\frac{\lambda\,b}{r}\,\extd r\Big)\tens v \cr
&& -\, \frac{2\,b}{r}\,v\wedge\nabla( v )
+\frac{3\,b\,\lambda}{r}\,v\wedge\nabla( \extd r) + \frac{\lambda\,b}{r}\,\extd r\wedge\nabla( v) \cr
&=& - 2\,b\,\frac{v\wedge\extd r}{r^2}\tens v 
+3\,b\,\lambda\,\frac{v\wedge\extd r}{r^2}\tens \extd r \cr
&& -\, \frac{2\,b}{r}\,v\wedge\nabla( v )
+\frac{3\,b\,\lambda}{r}\,v\wedge\nabla( \extd r) + \frac{\lambda\,b}{r}\,\extd r\wedge\nabla( v) \cr
&=& - 2\,b\,\frac{v\wedge \extd r}{r^2}\tens v 
+3\,b\,\lambda\,\frac{v\wedge\extd r}{r^2}\tens \extd r \cr
&& -\, \frac{2\,b}{r}\,v\wedge\Big(   -\frac{2}{r}\,v\tens \extd r 
-  \frac{b\,\lambda}{r}\,v\tens v  
+   \frac{\lambda}{r}\,\extd r\tens\extd r  \Big) \cr
&&+\, \frac{3\,b\,\lambda}{r}\,v\wedge \frac{2\,b}{r}\,v\tens v  + \frac{\lambda\,b}{r}\,\extd r\wedge\Big(   -  \frac{2}{r}\,v\tens \extd r  \Big) \cr
&=& - 2\,b\,\frac{v\wedge\extd r}{r^2}\tens v 
+3\,b\,\lambda\,\frac{v\wedge\extd r}{r^2}\tens \extd r \cr
&& -\, \frac{2\,b}{r}\,v\wedge\Big(   -\frac{2}{r}\,v\tens \extd r 
+   \frac{\lambda}{r}\,\extd r\tens\extd r  \Big)   - \frac{2\,\lambda\,b}{r^2}\,\extd r\wedge   v\tens \extd r  \cr
&=& - 2\,b\,\frac{v\wedge \extd r}{r^2}\tens v 
+5\,b\,\lambda\,\frac{v\wedge \extd r}{r^2}\tens \extd r \cr
&& +\, \frac{4\,b}{r^2}\,v\wedge v\tens \extd r 
 - \frac{2\,b\,\lambda}{r^2}\,v\wedge   \extd r\tens\extd r      \cr
 &=&- 2\,b\,\frac{v\wedge\extd r}{r^2}\tens v 
+7\,b\,\lambda\,\frac{v\wedge\extd r}{r^2}\tens \extd r    \ .
\end{eqnarray}

Finally, we start to compute Ricci as a contraction of the Riemann curvature. For this we have
\begin{eqnarray}  \label{vgdjyajx}
(\star\tens R)g &=& \overline{\extd r}\tens R(\extd r)+ b\, (\overline{v}\tens R(v) + \lambda (\overline{\extd r}\tens R(v)-\overline{v}\tens R(\extd r))) \cr
&=&  \overline{\extd r}\tens \Big(-2\,b\,\frac{v\wedge\extd r}{r^2}\tens v 
+7\,b\,\lambda\,\frac{v\wedge\extd r}{r^2}\tens \extd r\Big) \cr
&& +\, b\, \overline{v}\tens \Big(  \,2\,	\frac{v\wedge\extd r}{r^2}\tens \extd r 
+5\,b\,\lambda\, \frac{v\wedge\extd r}{r^2}\tens v\Big) \cr
&& +\, b\,  \lambda \, \overline{\extd r}\tens  \Big(  \,2\,	\frac{v\wedge\extd r}{r^2}\tens \extd r 
\Big)-b\,  \lambda \, \overline{v}\tens \Big(-2\,b\,\frac{v\wedge\extd r}{r^2}\tens v \Big)  \cr
&=&  \overline{\extd r}\tens v\wedge \extd r \tens\frac{b}{r^2} \Big(-2\, v 
+9\,\lambda\, \extd r\Big)  +  \overline{v}\tens v\wedge \extd r \tens \frac{b}{r^2} \Big( 2\,	\extd r 
+ 7\,b\,\lambda\, v\Big)  \ .
\end{eqnarray}

We will define the Ricci tensor from this by appling an interior product $\overline{\Omega^1}\tens_A \Omega^2\to\Omega^1$ which we will do as the
composition of a lift map $i:\Omega^2\to\Omega^1\tens_A \Omega^1$ with a sesquilinear pairing $\<\ ,\ \>:\overline{\Omega^1}\tens_A\Omega^1\to A$
given by inverting the sesquilinear metric (\ref{sesmetric2d}). In our case this 
comes out as \begin{eqnarray}
\<\overline{\extd r},\extd r\> &=& 1,\quad \<\overline{v},\extd r\>= \lambda,\quad 
\<\overline{\extd r},v\> = -\,\lambda,\quad \<\overline{v},v\> = b^{-1}  \ .
\end{eqnarray}
to order $\lambda$. 

Note that $\<\ ,\ \>$ is equivalent to working with the inverse $(\ ,\ )=\<\star(\ ),\ \>$ of $g$. For Ricci itself one can
clearly eliminate $\star$ from all these steps so that \cite{Ma:rsph,BegMa4,BegMa3}
\begin{equation}\label{riccireal} {\rm Ricci}=((\ ,\ )\tens\id)(\id\tens i\tens\id)((\id\tens R)(g)\end{equation}
if we wish. 

It remains to define $i:\Omega^2\to\Omega^1\tens_A \Omega^1$ to be a bimodule map and to obey $\wedge\,i=\id$. We do this with
3 parameters in the form
\begin{eqnarray*}
i(v\wedge\extd r)&=&\frac12\,v\tens \extd r- \frac12\,\extd r\tens v +\lambda\,\alpha\, \extd r\tens  \extd r  +  \, \lambda\,\beta(v\tens  \extd r+\extd r\tens  v)+\lambda\,\gamma\, v\tens  v\ .
\end{eqnarray*}
and calculate
\begin{eqnarray*}
(\<,\>\tens{\id})(\overline{\extd r}\tens i(v\wedge\extd r)) &=&- v/2-\lambda\,\extd r/2+\lambda\,\alpha\, \extd r
+\lambda\,\beta \, v\ ,\cr
(\<,\>\tens{\id})(\overline{v}\tens i(v\wedge\extd r)) &=& -\lambda\,v/2+b^{-1}\,\extd r/2
+\lambda\,\beta\,b^{-1}\, \extd r+\lambda\,\gamma\,b^{-1}\, v\ .
\end{eqnarray*}
Applying this to (\ref{vgdjyajx}) gives 
\begin{eqnarray*}
\mathrm{Ricci} &=&- \Big(v/2+\lambda\,\extd r/2-\lambda\,\alpha\, \extd r
-\lambda\,\beta \, v\Big)  \tens\frac{b}{r^2} \Big(-2\, v 
+9\,\lambda\, \extd r\Big) \cr
&&-\, \Big(\lambda\,v/2-b^{-1}\,\extd r/2
-\lambda\,\beta\,b^{-1}\, \extd r-\lambda\,\gamma\,b^{-1}\, v\Big) \tens \frac{b}{r^2} \Big( 2\,	\extd r 
+ 7\,b\,\lambda\, v\Big)\ ,
\end{eqnarray*}
which we rewrite as
\begin{eqnarray*}
r^2\,\mathrm{Ricci} &=&- b\, \Big(v/2+\lambda\,\extd r/2-\lambda\,\alpha\, \extd r
-\lambda\,\beta \, v\Big)  \tens \Big(-2\, v 
+9\,\lambda\, \extd r\Big) \cr
&&-\, \Big(\lambda\,b\,v/2-\extd r/2
-\lambda\,\beta\, \extd r-\lambda\,\gamma\, v\Big) \tens \Big( 2\,	\extd r 
+ 7\,b\,\lambda\, v\Big)\cr
&=& b\,v\tens v +\extd r\tens\extd r-2\,\lambda\,\beta(b\,v\tens v-\extd r\tens\extd r) \cr
&&+\, b\, \lambda\,(9/2-2\,\alpha) \, \extd r \tens v - \lambda\,(11\,b/2
-2\,\gamma)\,v \tens 	\extd r\ .
\end{eqnarray*}
Setting $\wedge\mathrm{Ricci} =0$ would give
\begin{eqnarray}\label{qsricci}
0\,=\,-\,\lambda\,b+b\, \lambda\,(9/2-2\,\alpha)+ \lambda\,(11\,b/2
-2\,\gamma)\ .
\end{eqnarray}

Finally, we impose $\star$-compatibility or `reality' in a suitable skew sense. For $i$ we note that $(v\wedge\extd r)^*=v\wedge\extd r$ in our conventions and in the general theory we want $i$ to anticommuta with $\star$. So what we require is that
\begin{eqnarray}\label{istar}
({\id}\tens\star)\,i(v\wedge\extd r)&=&+ \frac12\,v\tens \overline{\extd r}-\frac12\,\extd r\tens \overline{v^*} +\lambda\,\alpha\, \extd r\tens  \overline{\extd r} \cr
&& + \, \lambda\,\beta(v\tens  \overline{\extd r}+\extd r\tens  \overline{v})+\lambda\,\gamma\, v\tens  \overline{v} \cr
&=& \frac12\,v\tens \overline{\extd r}-\frac12\,\extd r\tens \overline{v} +\lambda\,(\alpha-1)\, \extd r\tens  \overline{\extd r} \cr
&& +  \, \lambda\,\beta(v\tens  \overline{\extd r}+\extd r\tens  \overline{v})+\lambda\,\gamma\, v\tens  \overline{v}\ 
\end{eqnarray}
should reverse sign under flip of the factors, conjugation of any coefficients and identification of the barred and unbarred elements. Note that  $\overline{v^*}=\overline{v-\lambda\,\extd r}=\overline{v}+\lambda\,\overline{\extd r}$ in this calculation. Inspecting the expression (\ref{istar}) we see that  `reality' or compatibility of $i$ with $\star$ corresponds to $\alpha,\beta,\gamma$ real.  

Putting this in, we compute
\begin{eqnarray}
r^2\,(\star\tens{\id})\mathrm{Ricci} 
&=& b\,\overline{v^*}\tens v +\overline{\extd r}\tens\extd r-2\,\lambda\,\beta(b\,\overline{v}\tens v-\overline{\extd r}\tens\extd r) \cr
&&+\, b\, \lambda\,(9/2-2\,\alpha) \, \overline{\extd r} \tens v - \lambda\,(11\,b/2
-2\,\gamma)\,\overline{v} \tens 	\extd r  \cr
&=& b\,\overline{v}\tens v +\overline{\extd r}\tens\extd r-2\,\lambda\,\beta(b\,\overline{v}\tens v-\overline{\extd r}\tens\extd r) \cr
&&+\, b\, \lambda\,(11/2-2\,\alpha) \, \overline{\extd r} \tens v - \lambda\,(11\,b/2
-2\,\gamma)\,\overline{v} \tens 	\extd r   \ 
\end{eqnarray}
and impose a requirement that  $\mathrm{Ricci}$ is `hermitian' in the same manner as for the metric $g$, as the equivalent  flip-invariance for the sesquilinear $(\star\tens\id){\rm Ricci}$.  This gives $\beta=0$ and (remembering that $\lambda$ is imaginary) $\gamma=b\,\alpha$. Our previous condition (\ref{qsricci}) for $\wedge\,\mathrm{Ricci} =0$ then gives $\alpha$ and we find
\begin{equation}\label{Ricci2d} \mathrm{Ricci} ={g\over r^{2}}\end{equation}
\begin{eqnarray}\label{i2d}
i(v\wedge\extd r)&=&\frac12\,v\tens \extd r-\frac12\,\extd r\tens v +{9\over 4}\lambda \, \extd r\tens  \extd r + {9\over 4} \lambda b\,   v\tens  v\ .
\end{eqnarray}
as the final answer. The lifting $i$ was a freedom in our theory but we have been led to a unique answer by the reality and quantum symmetry properties that we expect for {\rm Ricci}. 

From this the Ricci scalar defined as $S=(\ ,\ ){\rm Ricci}=\<\ ,\ \>(\star\tens\id)({\rm Ricci})$ via the metric inner product comes out as 
\begin{equation}\label{ricciscalar2d} S={2\over r^{2}}\end{equation}
to errors of order $\lambda^2$. If we define the Einstein tensor by its usual formula and
remembering that throughout the above we have been working to errors $O(\lambda^2)$, we conclude that
\begin{equation}\label{einst2d} {\rm Einstein}=O(\lambda^2).\end{equation}
So the classically 1+1 dimensional spacetime, which necessarily is a vacuum solution of Einstein's equations, has that same feature to linear order in $\lambda$.

\section{Exact noncommutative geometry of the 2D model}

Using the order $\lambda$ solutions of the preceding section as a model, we now exactly solve the noncommutative Riemannian geometry for our fixed metric (\ref{metric2d}). The computations now are much harder and done with the aid of Mathematica.  The first subsection analyses the connections  and finds, among other things, a unique Levi-Civita one that deforms the classical one (and extends the order $\lambda$ one already found). The second subsection looks at the Ricci tensor and finds that the quantum Levi-Civita connection obeys the noncommutative vacuum Einstein equations. 

\subsection{The quantum Levi-Civita connection}

Although we are interested in metric-compatible  torsion free  (or `Levi-Civita') connections, we also explore the moduli including torsion.
 We follow the same method as for 1st order and in particular we make an ansatz
\begin{equation}\label{conparama} \nabla\extd r= {1\over r} (\alpha v\tens v+ \beta v\tens\extd r+ \gamma\extd r\tens v+\delta \extd r\tens\extd r)\end{equation}
\begin{equation}\label{conparamb} \nabla v= {1\over r} (\alpha' v\tens v+ \beta' v\tens\extd r+ \gamma'\extd r\tens v+\delta' \extd r\tens\extd r)\end{equation}
inspired by our order $\lambda$ solution, i.e. keeping the same form but with coefficients that are functions of $\lambda$ but not of $t,r$. 
The torsion $T=\wedge\nabla-\extd$ (say), using $v^2=\lambda v\wedge\extd r$ and $\extd v=-{2 \over r}v \wedge \extd r$, is
\[ T(\extd r)={1\over r}(\lambda\alpha+\beta-\gamma)v\wedge\extd r,\quad T(v)={1\over r}(\lambda\alpha'+\beta'-\gamma'+2)v\wedge\extd r\]
and the braiding by the same formula as before comes out as:
\[\sigma(v\tens v)=(1+\lambda\alpha')v\tens v+\lambda\beta'v\tens \extd r+\lambda\gamma'\extd r\tens v+\lambda\delta' \extd r\tens\extd r\]
\[ \sigma(\extd r\tens v)=(1+\lambda\beta)v\tens\extd r+\lambda\alpha v\tens v+ \lambda\gamma\extd r\tens v+ \lambda\delta \extd r\tens\extd r\]
\[ \sigma(x\tens\extd r)=\extd r\tens x\]

Then $*$-preserving comes out as
\[ \alpha-\overline\alpha=\lambda(\overline\alpha\alpha'+\alpha\overline\beta)-\lambda^2|\alpha|^2\]
\[ \beta-\overline\beta=\lambda(|\beta|^2+\overline\alpha(\beta'-1))-\lambda^2\overline\alpha\beta\]
\[ \gamma-\overline{\gamma}=\lambda(\overline{\beta}\gamma+\overline{\alpha}(\gamma'-1))-\lambda^2\overline{\alpha}\gamma\]
\[ \delta-\overline\delta=\lambda(\overline\alpha\delta'+\overline\beta(\delta-1)-\overline{\gamma})-\lambda^2\overline\alpha(\delta-1)\]
from $\nabla \extd r$ and 
\[ \alpha'-\overline{\alpha'}=\lambda(\alpha(\overline{\beta'}+1)+|\alpha'|^2)-\lambda^2\overline{\alpha'}\alpha\]
\[ \beta'-\overline{\beta'}=\lambda(\beta(\overline{\beta'}+1)+\overline{\alpha'}(\beta'-1))-\lambda^2\overline{\alpha'}\beta\]
\[ \gamma'-\overline{\gamma'}=\lambda(\gamma(\overline{\beta'}+1)+\overline{\alpha'}(\gamma'-1))-\lambda^2\overline{\alpha'}\gamma\]
\[ \delta'-\overline{\delta'}=\lambda(\overline{\alpha'}\delta'+(\overline{\beta'}+1)(\delta-1)-(\overline{\gamma'}-1))-\lambda^2\overline{\alpha'}(\delta-1)\]
from $\nabla v$. Using Mathematica and assuming $\alpha\ne 0$, this system is solved by an arbitrary choice of $\alpha,\beta,\gamma,\delta$, say, and
\begin{eqnarray}\label{ap} \alpha'&=&{1\over\lambda\bar\alpha}(\alpha-\bar\alpha-\lambda\bar\beta \alpha+\lambda^2|\alpha|^2),\\
\label{bp} \beta'&=&{1\over\lambda\bar\alpha}(\beta-\bar\beta+\lambda(\bar\alpha-|\beta|^2)+\lambda^2\bar\alpha\beta)\\ 
\label{cp} \gamma'&=&{1\over\lambda\bar\alpha}(\gamma-\bar\gamma+\lambda(\bar\alpha-\bar\beta\gamma)+\lambda^2\bar\alpha\gamma)\\
\label{dp} \delta'&=&{1\over\lambda\bar\alpha}(\delta-\bar\delta+\lambda(\bar\gamma-\bar\beta(\delta-1))+\lambda^2\bar\alpha(\delta-1))\end{eqnarray}

Next, we have 
\[ \nabla\overline{\extd r}= {1\over r} (\bar\alpha \bar v\tens v+ (\bar\gamma-\lambda\bar\alpha)\bar v\tens\extd r+ \bar\beta\overline{\extd r}\tens v+(\bar\delta-\lambda\bar\beta)\overline{ \extd r}\tens\extd r)\]
\[ \nabla \overline{v}= {1\over r} (\overline{\alpha'} \bar v\tens v+ (\overline{\gamma'}-\lambda\overline{\alpha'})\bar v\tens\extd r+ \overline{\beta'}\, \overline{\extd r}\tens v+(\overline{\delta' }-\lambda\overline{\beta'})\overline{\extd r}\tens\extd r)\]
and compute *-metric compatibility by acting with $\nabla$ on $(\star\tens\id)(g)$ as a derivation. There are 32 terms which we regroup as 4 terms for each of the 8 basis elements $\extd r\tens\extd r\tens\extd r,\cdots,v\tens v\tens v$. They are not all independent and we obtain the following  6 equations
 \begin{eqnarray}\label{mc1} \alpha'+\overline{\alpha'}&=&\lambda(\alpha-\bar\alpha)\\ 
\label{mc2} \beta+\bar\beta&=&-b\lambda(\beta'-\overline{\beta'})\\ 
\label{mc3}  \alpha+\lambda b\bar\beta&=&-b(\overline{\beta'}+\lambda\alpha')\\
\label{mc4}  \gamma'+\overline{\gamma'}&=&\lambda(\gamma-\bar\gamma+\overline{\alpha'}+\lambda\bar\alpha)\\
\label{mc5}  \delta+\overline{\delta}+b\lambda (\delta'-\overline{\delta'})&=&\lambda (\overline{\beta}-\lambda b\overline{\beta'})\\
\label{mc6} \gamma+\lambda b\gamma'+b(\overline{\delta'}+\lambda\bar\delta)&=&\lambda b(\overline{\beta'}+\lambda\bar\beta).\end{eqnarray}
Before studying these equations we note that the third one minus its conjugate implies, for $1+b\lambda^2\ne 0$, that 
\begin{equation}\label{mc3new}\alpha-\bar\alpha=b(\beta'-\overline{\beta'}).\end{equation}
This in conjunction with the second equation tells us that
\begin{equation}\label{extrametric} \beta+\bar\beta=-\lambda(\alpha-\bar\alpha)\end{equation}
which will be useful. We assume throughout that $b,\lambda,1+b\lambda^2\ne 0$ and we note that the Lorentzian case has $b<0$. One can show if $1+{b\lambda^2\over 2}\ne 0$ that $\alpha=0$ implies that the entire solution is zero, so we
exclude $\alpha=0$ in our analysis and put it back in by hand in our final results.  One can also show from (\ref{mc1})-(\ref{mc6}) that $T(v)$ is determined in a simple way from $T=T(\extd r)$,
\begin{equation}\label{Tv}  T(v)= \lambda T+ {T-\overline{T}\over\lambda\bar\alpha}-{\bar\beta\over\bar\alpha}T.  \end{equation}
 Hence we need only focus on $T(\extd r)$ and if this vanishes then so does the whole torsion. 

The most relevant  solutions turn out to be `real' in the sense:  
\begin{equation}\label{realconpar}\alpha,\delta,\beta',\gamma'\in \R,\quad \alpha',\delta',\beta,\gamma\in \imath\R\end{equation} 
as we shall see. It is evident that if the unprimed variables obey (\ref{realconpar}) then they all do. More surprisingly our main class of exact solutions turn out to be characterised better by a novel  property 
\begin{equation}\label{gammaalph} \gamma=-{\lambda \alpha\over 2},\quad  \delta=-{\lambda \beta\over 2},\quad  \gamma'=-{\lambda \alpha'\over 2},\quad  \delta'=-{\lambda \beta'\over 2}\end{equation}
which means of course that
\[ \nabla\extd r= {1\over r}  (v-{\lambda\extd r\over 2})\tens (\alpha v+ \beta\extd r),\quad \nabla v= {1\over r}  (v-{\lambda\extd r\over 2})\tens (\alpha' v+ \beta'\extd r)\]
so we say such connections are `decomposable'.  It is easy to see that if the unprimed variables obey (\ref{gammaalph}) then they all do.

\begin{proposition} \label{moduli}  The space of  `real' $*$-preserving metric compatible connections consists of (1) a conic
\[ \beta(\beta+ {2\over\lambda})+\alpha(1+\lambda\beta)+{\alpha^2\over b}(1+b\lambda^2)=0;\quad \gamma=-{\lambda\over 2}\alpha,\quad \delta=-{\lambda\over 2}\beta\]
where $\alpha$ is real and $\beta$ is imaginary and (2) a line $\R$ for a real parameter $\delta$, with
\[  \alpha={b\over 1+b\lambda^2},\quad \beta=-{2\over\lambda},\quad\gamma={2+{b\lambda^2}\over 2\lambda( 1+b\lambda^2)}-{\delta\over\lambda},\]
and passing through the conic at $\delta=1$.  

The full space of $*$-preserving metric compatible connections consists in case (1) of  a line $\R$ for the imaginary value $\delta-\bar\delta$ (we now allow $\delta$ to be complex),  at each point of the conic, with 
\[   \gamma=-{\alpha\over 2}(\lambda-{\delta-\bar\delta\over \beta}),\quad \delta+\bar\delta=-\beta\lambda\]
and in case (2) of $\C\times\R$ for a complex parameter $\delta$ and a free real parameter $\gamma+\bar\gamma$,  with
\[ \lambda(\gamma-\bar\gamma)+ \delta+\bar\delta= {2+ b\lambda^2\over 1+b\lambda^2}.\]
\end{proposition}
\begin{figure}
\includegraphics[scale=.42]{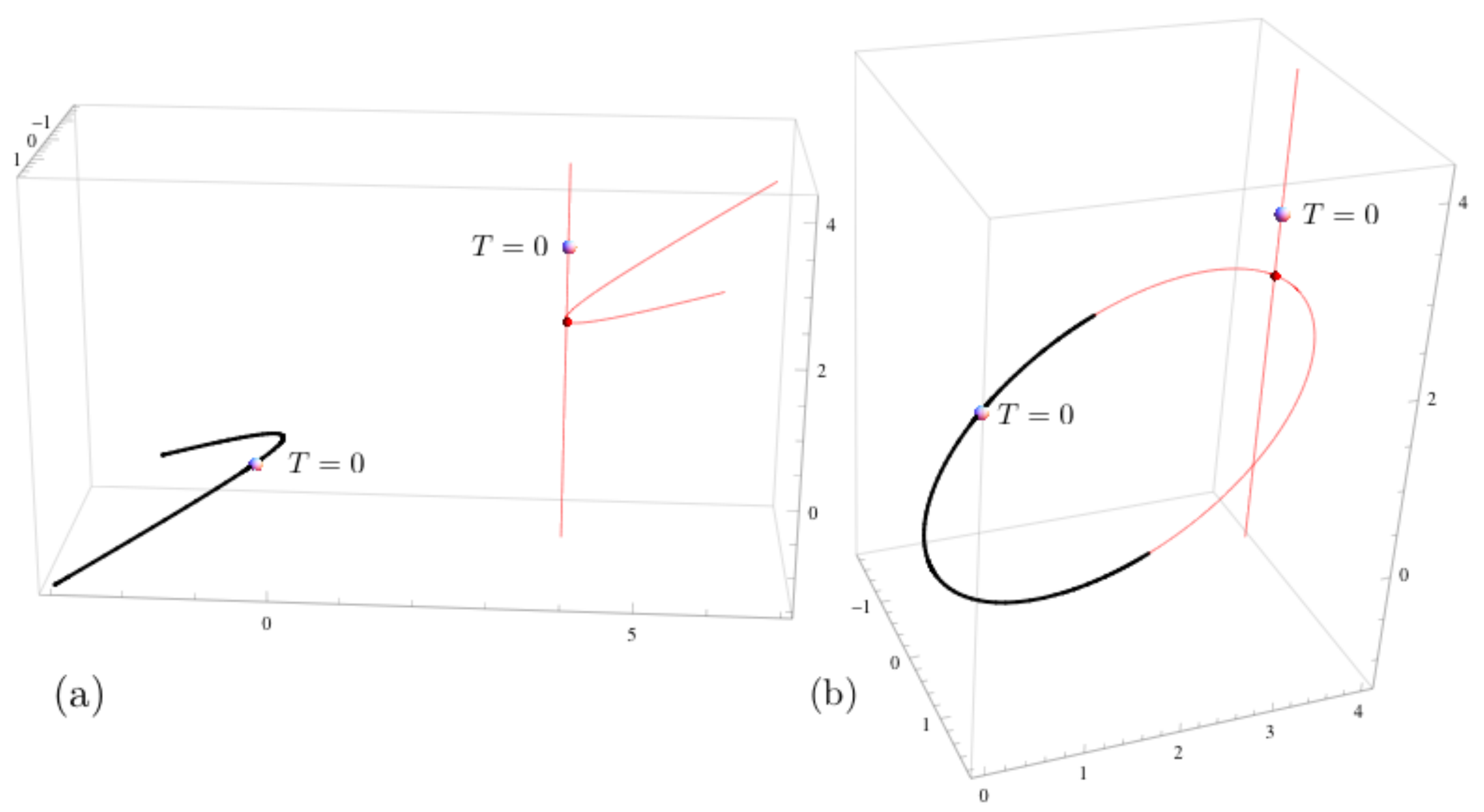}
\caption{Space of `real' $*$-preserving metric compatible connections at (a) $b>0$ and (b) $b<0$. The bold black curves are the branch of the square-root with classical limit as $\lambda\to 0$, the other part of the conic and intersecting line are non-perturbative. In each case there is a unique torsion free or `Levi-Civita' point as marked.  The axes are $\alpha,-\imath\beta$ horizontally and $\delta$ vertically.}
\end{figure}
\proof We state only the unprimed variables, with the primed ones being determined from (\ref{ap})-(\ref{dp}) so as to solve the $*$-preservation condition. In case (1) this means
\[ \alpha'=\lambda\alpha+\beta,\quad \beta'=-{\alpha\over b}(1+b\lambda^2),\quad \gamma'=-{\alpha'\over 2}(\lambda-{\delta-\bar\delta\over \beta}),\quad \delta'=-{\beta'\over 2}(\lambda-{\delta-\bar\delta\over \beta})\]
and in case (2) it means
\[ \alpha'=-{2+b\lambda^2\over\lambda(1+b\lambda^2)},\quad \beta'=-1,\quad \gamma'={4+{3b\lambda^2}\over 2( 1+b\lambda^2)}-({\delta+\bar\delta\over 2})-({2+b\lambda^2\over  b\lambda})({\gamma+\bar\gamma\over 2}),\]
\[ \delta'={2+b\lambda^2-(\delta+\bar\delta)\over  2 b\lambda}+{\lambda(\delta-\bar\delta)\over 2}+{(\gamma+\bar\gamma)\over 2b}(1+b\lambda^2).\]

We first show that $\alpha$ is necessarily real and $\beta$ imaginary. Let $z=\alpha-\bar\alpha$ and suppose that $z\ne 0$. We solve the $*$-preservation condition by defining the prime variables from the unprimed ones according to (\ref{ap})-(\ref{dp}). Then
\begin{eqnarray*} \alpha'+\overline{\alpha'}&=&{\alpha(1-\lambda\bar\beta)\over\lambda\bar\alpha}-{\bar\alpha(1+\lambda\beta)\over\lambda\alpha}+\lambda z.
\end{eqnarray*}
Hence (\ref{mc1}) means $\alpha^2(1-\lambda\bar\beta)= \bar\alpha^2(1+\lambda\beta)$. Using (\ref{extrametric}) we have $(\alpha^2-\bar\alpha^2)(1+\lambda\beta)=-\lambda^2\alpha^2z$ hence
\begin{equation}\label{notrealA}(\alpha+\bar\alpha)(1+\lambda\beta)=-\lambda^2\alpha^2.\end{equation}
In this case we see that $\alpha\ne 0$ cannot be purely imaginary either and we obtain $\beta$ as a function of $\alpha$. Similarly 
\begin{eqnarray*}b(\beta'-\bar\beta')&=& b\lambda(\beta+\bar\beta)+b\left({1\over\bar\alpha}-{1\over\alpha}\right)\left({\beta-\bar\beta\over\lambda}-|\beta|^2\right). 
\end{eqnarray*} 
Hence (\ref{mc3new}) and (\ref{extrametric}) tells us that 
\begin{eqnarray}\label{notrealB} (1+b\lambda^2)|\alpha|^2=b \left({\beta-\bar\beta\over\lambda}-|\beta|^2\right). \end{eqnarray}
Finally, (\ref{mc3}) means
\[ \alpha+\lambda b\bar\beta=-b\left({\beta-\bar\beta\over\lambda\alpha}-{|\beta|^2\over\alpha}-\lambda\bar\beta+{\alpha\over\bar\alpha}(1-\lambda\bar\beta)+\lambda^2\alpha\right)\]
or
\begin{equation}\label{notrealC} (1+b\lambda^2)|\alpha|^2=-b\alpha(1-\lambda\bar \beta)-{b\bar\alpha\over\alpha}\left({\beta-\bar\beta\over\lambda}-|\beta|^2\right).\end{equation}
Putting in (\ref{notrealB}) and then (\ref{notrealA}), we have
\[ - b\alpha(1-\lambda\bar\beta)=(1+b\lambda^2)|\alpha|^2(1+{\bar\alpha\over\alpha})=(1+b\lambda^2)|\alpha|^2{\alpha+\bar\alpha\over\alpha}=-(1+b\lambda^2){\lambda^2|\alpha|^2\alpha\over  1+\lambda\beta} \]
Cancelling $\alpha$ we conclude that 
\[ \lambda^2(1+b\lambda^2)|\alpha|^2=b(1-\lambda\bar\beta)(1+\lambda\beta)=b+b\lambda^2\left({\beta-\bar\beta\over\lambda}-|\beta|^2\right)\]
which contradicts (\ref{notrealB}) as $b,\lambda\ne 0$. 

Hence $\alpha$ is necessarily real. In this case (\ref{extrametric}), (\ref{mc3new}) and (\ref{mc1}) tell us that $\alpha',\beta$ are imaginary and $\beta'$ is real and indeed
\[ \alpha'=\beta+\lambda\alpha,\quad \beta'={\beta\over\alpha}(\beta+{2\over\lambda})+1+\lambda\beta.\]
At this point
(\ref{mc1}) and (\ref{mc2}) are empty while (\ref{mc3}) becomes $\alpha-b\lambda\beta=-b\beta'-b\lambda(\beta+\lambda\alpha)$ or $\beta'$ as stated. Comparing with (\ref{bp}) r from (\ref{notrealC}) (which still holds when $\alpha$ is real) we see that
\begin{equation}\label{quadric} \beta(\beta+ {2\over\lambda})+\alpha(1+\lambda\beta)+{\alpha^2\over b}(1+b\lambda^2)=0\end{equation}
which is our conic and which also allows us to give $\beta'$ as a function of $\alpha$. Although we excluded $\alpha=0$ from our analysis we can put this back now and in other final answers. This is the content of (\ref{mc1})-(\ref{mc3}). 

Next, (\ref{cp})-(\ref{dp}) become
\[\gamma'=1+\lambda\gamma+{\gamma(1+\lambda\beta)-\bar\gamma\over\lambda\alpha},\quad \delta'=\lambda(\delta-1)+{\delta(1+\lambda\beta)-\bar\delta+\lambda(\bar\gamma-\beta)\over\lambda\alpha}\]
from the first of which (\ref{mc4}) becomes
\[ \gamma'+\overline{\gamma'}=2+(\lambda+{ 2+\lambda\beta\over\lambda\alpha})(\gamma-\bar\gamma)=\lambda(\gamma-\bar\gamma)-\lambda\beta\]
or 
\[ (2+\lambda\beta)(1 + {\gamma-\bar\gamma\over\lambda\alpha})=0.\]
Hence there are two ways to satisfy (\ref{mc4}), being the two stated routes (1) and (2). In the case (2) we solve the quadratic (\ref{quadric}) for $\alpha$.

Next, (\ref{mc5})  becomes 
\[ \delta+\bar\delta+b\lambda(\delta'-\overline{\delta'})+\lambda\beta-\lambda^2{\alpha (1+b\lambda^2)}=0\]
where 
\[ \delta'-\overline{\delta'}=-{\gamma-\bar\gamma\over\alpha}+({\beta\over\alpha}+\lambda)(\delta+\bar\delta-2).\]
In case (2) we use $2+\lambda\beta=0$ and $\alpha(1+b\lambda^2)=b$ to simplify this requirement down to 
the stated relation between $\gamma-\bar\gamma$ and $\delta+\bar\delta$.  In case (1) we use (\ref{quadric}) to simply the
requirement down to $\delta+\bar\delta=-\lambda\beta$. 

Finally,  (\ref{mc6}) becomes
\[ \gamma(1+\lambda^2b + {b(2+\lambda\beta)\over\alpha})-\bar\gamma{b\over\alpha}+{b\over\lambda\alpha}(\delta-\bar\delta(1+\lambda\beta))+{b\over\alpha}\beta+b\lambda (2+\lambda\beta)+\lambda\alpha(1+b\lambda^2)=0\]
For case (2) this gives the same result as for (\ref{mc5}) while for case (1) we add the equation to its conjugate which then simplifies down to 
\[ (\alpha(1+b\lambda^2)+b(1+\lambda\beta))(\gamma+\bar\gamma)+b(\beta+{2\over\lambda})(\delta-\bar\delta)=0.\]
On using (\ref{quadric}) again, this gives $\gamma+\bar\gamma={\alpha\over \beta}(\delta-\bar\delta)$. Combining with $\gamma-\bar\gamma=-\lambda\alpha$ in case (1)  gives $\gamma$ as stated for this case.

The remaining values of the primed variables are then determined from the above. The intersection of the two parts of the moduli space requires $\delta+\bar\delta=2$, the imaginary part of $\delta$ remains a free parameter (so the intersection is a real line). Here $\gamma$ is determined as
\[ \gamma=-{b\lambda\over 2(1+b\lambda^2)}(1-{\delta-\bar\delta\over 2})\]
according to the above. Among `real' connections we have a unique point of intersection, with $\delta=1$.  \endproof

We note that in case (1) we can regard $\alpha$ as a free real parameter within a certain range and solve for $\beta$ as 
\begin{equation}\label{betaroot} \beta=-{1\over 2\lambda}\left(2+\alpha \lambda^2\mp\sqrt{4-{\alpha^2 \lambda^2\over b}\left(4 +3 b \lambda^2\right)}\right)\end{equation}
where of the two branches only the (-) has a classical limit. The other side of the conic has no classical analogue as it blows up as $\lambda\to 0$. 

Also note that  if we have a solution of our equations  of type (1) above then
\begin{equation}\label{gauge1} \gamma\to \gamma+\alpha\Delta,\quad\delta\to \delta+\beta\Delta\quad \gamma'\to\gamma'+(\beta+\lambda\alpha)\Delta,\quad \delta'\to\delta'+((1+\lambda\beta)+{\beta\over\alpha}(\beta+{2\over\lambda}))\Delta\end{equation}
is another solution for any real $\Delta$. Similarly if we have a solution of type (2) above then
\begin{equation}\label{gauge2}\delta \to \delta+\Delta_1,\quad \delta'\to \delta'+\lambda\Delta_1,\quad \gamma\to \gamma+\Delta_2,\quad \gamma'\to \gamma'- {2+b\lambda^2\over\lambda b}\Delta_2\end{equation}
is another solution for any imaginary $\Delta_1$ and any real $\Delta_2$.  In this way any solution can be transformed to a unique `real' one in the same family, in the sense of (\ref{realconpar}). We can therefore focus on `real' solutions. 

\begin{corollary}\label{levi}  The *-preserving metric compatible connections with zero torsion are either: (1) a unique one with a classical limit,
\[ \alpha={8b\over 4+7 b\lambda^2},\quad \beta=-{12 b\lambda\over 4+7 b\lambda^2},\quad 
 \alpha'=-{4b\lambda\over 4+7 b\lambda^2},\quad  \beta'=-2{(1+b\lambda^2)\over 4+7 b\lambda^2} .\]
This is both `real' and decomposable in the sense (\ref{realconpar}), (\ref{gammaalph}). 
(2) a unique one up to (\ref{gauge2}) without classical limit,
\[ \alpha={b\over  1+b\lambda^2},\quad\beta=-{2\over\lambda},\quad \gamma=-{2+{b\lambda^2}\over \lambda( 1+b\lambda^2)},\quad\delta={3(2+ b\lambda^2)\over 2(1+b\lambda^2)}\]
\[ \alpha'=-{2+{b\lambda^2}\over\lambda( 1+b\lambda^2)},\quad \beta'=-1,\quad \gamma'=-{1\over 1+b\lambda^2},\quad \delta'={2+{b\lambda^2}\over 2 b\lambda}.\]
\end{corollary}
\proof We impose $T(\extd r)=0$, i.e. we set  $\beta=\gamma-\lambda\alpha$ (and find that $T(v)=0$) hence $\alpha'=\gamma$ in both cases. It also follows that $\gamma$ must be imaginary. In case (1) this fixes any freedom and we have to have $\delta$ real and $\gamma=-{\lambda\over 2}\alpha$. Hence we need ${3\lambda\over 2}\alpha+\beta=0$ which fixes $\alpha$ as stated. We also need to take the (-) branch of the square-root.  In case (2) since $\alpha,\beta$ are fixed we have a unique $\gamma=\beta+\lambda\alpha$ as stated. This then fixes $\delta+\bar\delta=3 (2+\lambda^2)/(1+b\lambda^2)$. The imaginary part of $\delta$ is not fixed but we can take $\delta$ to be real if we want, as stated.  \endproof

The two Levi-Civita points are shown in Figure~4. We see that only one of them has a classical limit so we recover classical uniqueness, but that a unique other connection is possible at the nonperturbative level.  More generally the torsions are
\begin{equation}\label{tor1} T(\extd r)=\left({\lambda\alpha\over 2}+\beta - {\alpha(\delta-\bar\delta)\over 2\beta}\right){v\wedge\extd r\over r} \end{equation}
in case (1) and 
\begin{equation}\label{tor2} T(\extd r)=\left({(\delta+\bar\delta)\over 2\lambda}-{3(2+{b\lambda^2 })\over 2 \lambda (1+b\lambda^2)}-{(\gamma+\bar\gamma)\over 2}\right){v\wedge\extd r\over r} \end{equation}
in case (2), from which the zero points are again clear. For the torsion to be real we should stick to the case where $\delta$ is real in case (1) and $\gamma$ is imaginary in case 2, which is the case for the `real' moduli. The connection (1) is also the only one of the two to be decomposable.

\begin{corollary}\label{A1} $*$-preserving metric-compatible decomposable connections  are precisely the real conic in Proposition~\ref{moduli}. \end{corollary}
\proof  Decomposable requires that we fix $\gamma=-{\lambda\over 2}\alpha$ which then forces is in case (1) of the moduli space in Proposition~\ref{moduli} to have $\delta$ to be real and hence uniquely determined as $\delta=-\lambda\beta/2$, as also required for a decomposable connection. In case (2) the decomposability fixes $\delta=1$ which takes us back to a point of case (1).  \endproof

We see that `real' moduli space of $*$-preserving metric compatible connections is connected and broadly similar to the classical situation with the torsion being more or less free to prescribe but with some nonlinearities and additional branches  due to the finite $\lambda$. Intersecting the non-classical part of the cone we also have a nonclassical line allowing us again to prescribe the torsion and we have the possibility of complex extensions by transformations (\ref{gauge1})-(\ref{gauge2}). 

Finally, we settle an issue which has no classical analogue: one could ask when the associated $\sigma$ of the bimodule connection obeys the braid relations.  If $\lambda=0$ then this holds for all connections with $\sigma$ the flip map. However, if $\lambda\ne 0$ and $\alpha\ne 0$ (say) then we find that connections of the form (\ref{conparama})-(\ref{conparamb}) obey the braid relations if $\alpha,\beta,\gamma$ are otherwise free (so a 3 complex dimensional parameter space) and
\[\delta=\frac{\beta \gamma}{\alpha},\quad\alpha'=-{1+\lambda\beta\over\lambda},\quad
\beta'=- \frac{\beta (1+\lambda\beta )}{\alpha \lambda},\]
\[\gamma'=-\frac{\gamma (1+ \lambda \beta)}{\alpha  \lambda},\quad 
   \delta'=-\frac{\beta \gamma (1+ \lambda\beta)}{\alpha^2 \lambda}.\]
These are non-perturbative and have no intersection with `real' $*$-preserving connections obeying (\ref{ap})-({\ref{dp}) nor with the $*$-preserving metric compatible connections in Proposition~\ref{moduli}, at least for generic $b\lambda^2$.
  
\subsection{Ricci curvature}
Write the Riemann curvature in the following form,
\begin{eqnarray}
R(\extd r)&=&-\,\frac{1}{r^2}\, v\wedge\extd r\tens(c_1\,v+c_2\,\extd r)\cr
R(v)&=& -
\frac{1}{r^2}\, v\wedge\extd r\tens(c_3\,v+c_4\,\extd r)\ .
\end{eqnarray}
where $c_1,c_2,c_3,c_4$ are calculated from the coefficients in (\ref{conparama})-(\ref{conparamb}),
\begin{eqnarray}\label{ciformula}
c_1 &=& \alpha  \left(\lambda  \alpha '+\beta  \lambda +\gamma '-\delta +1\right)+\gamma 
\left(\beta -\alpha '\right)\ ,\cr
 c_2 &=& \alpha  \lambda  \beta '+\alpha  \delta '+\beta ^2 \lambda -\gamma  \beta '+\beta\ ,\cr
 c_3 &=& \lambda  \left(\alpha '\right)^2+\alpha '+\beta ' (\alpha  \lambda +\gamma )-\alpha 
   \delta '\ ,\cr
 c_4 &=& \beta ' \left(\lambda  \alpha '+\beta  \lambda -\gamma '+\delta +1\right)+\delta '
   \left(\alpha '-\beta \right)\ .
\end{eqnarray}
If we impose the reality constraints (\ref{realconpar}) then we find 
\begin{equation}\label{realcurv}c_1,c_4\in\mathbb{R},\quad c_2,c_3\in\mathrm{i}\,\mathbb{R}\end{equation}
and whenever this happens we say that the curvature coefficients are `real'.

Following the same line and methods as at order $\lambda$, we again define Ricci as
\begin{equation*}\label{riccirealb} {\rm Ricci}=((\ ,\ )\tens\id)(\id\tens i\tens\id)((\id\tens R)(g)\end{equation*}
via a lifting map $i:\Omega^2\to \Omega^1\tens\Omega^1$ and our next result is that this map is uniquely determined by the required symmetry and reality properties of {\rm Ricci}. This is not quite as in classical geometry, where $i$ is defined independently, but the upshot is the same. 
We work always with our fixed metric (\ref{metric2d}). As with our analysis for the connection, we assume a linear form where $i(v\wedge\extd r)$ is a linear combination
of tensor products of $v,\extd r$.

\begin{proposition}\label{riccici} Let $c_i$ be  `real' curvature coefficients for the Riemann tensor of a connection. There is a unique skew-hermitian lift $i$ such that the {\rm Ricci} tensor has the same `hermitian' and quantum symmetry properties as the metric. In this case  
\begin{eqnarray*}
\mathrm{Ricci} &=&-\frac{\left(1+b \lambda ^2\right) (c_2 c_3-c_1 c_4)}{2 r^2 
   (c_4-\lambda  (c_2-c_3+c_1 \lambda ))}\Big(v^*\tens v +  \lambda( \extd r\tens v -\, v^*\tens \extd r) \cr
&&\quad\quad\quad\quad +    \frac{\lambda  \left(\left(1+b \lambda ^2\right) (c_1 \lambda
   -c_3)+c_2 \left(2+b \lambda
   ^2\right)\right)-c_4}{c_1+c_3 b \lambda }\,\extd r\tens \extd r\Big)\ .
\end{eqnarray*}
We assume that the $c_i$ are such that the denominators do not vanish.
\end{proposition}
\proof We let
\begin{eqnarray}
i(v\wedge\extd r) &=& n_1\,v\tens v + (n_2-\lambda\,n_1)\,v\tens \extd r \cr &&  +\, 
(n_3-\lambda\,n_1)\,\extd r\tens v +
(n_4-\lambda\,n_3-\lambda\,n_2+\lambda^2\,n_1)\,\extd r\tens \extd r \ ,
\end{eqnarray}
for some numerical coefficients $n_i$, or equivalently, 
\begin{eqnarray}
(\star\tens\star)\,i(v\wedge\extd r) &=& n_1\,\overline{v}\tens \overline{v} + n_2\,\overline{v}\tens \overline{\extd r} +
n_3\,\overline{\extd r}\tens \overline{v} +
n_4\,\overline{\extd r}\tens \overline{\extd r} \ ,
\end{eqnarray}
which we will need later. The condition that  $\wedge i=\id$ is that $n_2-n_3+\lambda\,n_1=1$. 
From the equation
\begin{eqnarray}
(\star\tens\id)\,i(v\wedge\extd r) &=& n_1\,\overline{v}\tens v + (n_2-\lambda\,n_1)\,\overline{v}\tens \extd r \cr &&  +\, 
n_3\,\overline{\extd r}\tens v +
(n_4-\lambda\,n_3)\,\overline{\extd r}\tens \extd r \ ,
\end{eqnarray}
we get the condition for $i$ to be `skew-hermitian' in the same sense as we required at order $\lambda$. This comes out as $n_1$ and $n_4-\lambda\,n_3$ imaginary and $n_2-\lambda\,n_1=-\,n_3^*$. 

Next we will work with the sesquilinear inner product $\<\ ,\  \>=(\ ,\star^{-1}(\ ))$
\begin{eqnarray*}
\<v,\overline{v}\>\,=\,\frac{1}{b\,(1+\lambda^2b)}\ ,\ \<v,\overline{\extd r}\>\,=\,-\<\extd r,\overline{v}\>\,=\,\frac{\lambda}{1+\lambda^2b}\ ,\ \<\extd r,\overline{\extd r}\>\,=\,\frac{1}{1+\lambda^2b}\ .
\end{eqnarray*}
equivalent to the ordinary inverse of the metric,
\begin{eqnarray}\label{roundinner}
(v^*,v)={1\over b},\quad (v^*,\extd r)=0=(\extd r, v),\quad (\extd r,\extd r)={1\over 1+ b\lambda^2}.
\end{eqnarray}
We also prefer to write the metric as
\begin{equation}\label{metric2dvv} 
 g=((1+b\,\lambda^2)\extd r-b\,\lambda\,v)\tens\extd r+ b\,v\tens v\ .
\end{equation}

We are then ready to compute
\begin{eqnarray*}
-\,r^2\,(\star\tens\id)\mathrm{Ricci} &=&-\,r^2\,(\<,\>\tens\id\tens\id)(\id\tens (\star\tens\star)\,i\tens\id)(\id\tens R)g \cr
&=& \<(1+b\,\lambda^2)\extd r-b\,\lambda\,v,\overline{v}\>\, (n_1\,\overline{v}+n_2\,\overline{\extd r})\tens (c_1\,v+c_2\,\extd r) \cr
&&+\,  \<(1+b\,\lambda^2)\extd r-b\,\lambda\,v,\overline{\extd r}\>\, (n_3\,\overline{v}+n_4\,\overline{\extd r})\tens (c_1\,v+c_2\,\extd r) \cr
&& +\,  \<b\,v,\overline{v}\>\, (n_1\,\overline{v}+n_2\,\overline{\extd r})\tens (c_3\,v+c_4\,\extd r) \cr
&&+\,  \<b\,v,\overline{\extd r}\>\, (n_3\,\overline{v}+n_4\,\overline{\extd r})\tens (c_3\,v+c_4\,\extd r) \cr
&=&\big(-\,\lambda\,(2+\lambda^2\,b)\, (n_1\,\overline{v}+n_2\,\overline{\extd r})\tens (c_1\,v+c_2\,\extd r) \cr
&&+\,  (n_3\,\overline{v}+n_4\,\overline{\extd r})\tens (c_1\,v+c_2\,\extd r) \cr
&& +\,  (n_1\,\overline{v}+n_2\,\overline{\extd r})\tens (c_3\,v+c_4\,\extd r) \cr
&&+\, \lambda\,b\, (n_3\,\overline{v}+n_4\,\overline{\extd r})\tens (c_3\,v+c_4\,\extd r)\big)/(1+\lambda^2\,b)\cr
&=& (x\,\lambda\,\overline{v}+y\,\overline{\extd r})\tens (c_3\,v+c_4\,\extd r) +
(p\,\overline{v}+\lambda\,q\,\overline{\extd r})\tens (c_1\,v+c_2\,\extd r) \  \cr
&=& (\lambda\,x\,c_3+p\,c_1)\,\overline{v}\tens v +  (y\,c_3+\lambda\,q\,c_1)\,\overline{\extd r}\tens v \cr
&&+\,  (\lambda\,x\,c_4+p\,c_2)\,\overline{v}\tens \extd r+ (y\,c_4+\lambda\,q\,c_2)\,\overline{\extd r}\tens \extd r\ .
\end{eqnarray*}
where for short we have put
\begin{eqnarray*}
\lambda\,x &=& (n_1+\lambda\,b\,n_3)/(1+\lambda^2\,b)\ ,  \cr
y &=& (n_2+\lambda\,b\,n_4)/(1+\lambda^2\,b)\ ,  \cr
p &=& \big(-\,\lambda\,(2+\lambda^2\,b)\, n_1+n_3   \big)/(1+\lambda^2\,b)\ ,  \cr
\lambda\,q &=& \big(-\,\lambda\,(2+\lambda^2\,b)\, n_2+n_4   \big)/(1+\lambda^2\,b)\ .
\end{eqnarray*}

From this we also get
\begin{eqnarray*}
-\,r^2\,\mathrm{Ricci} &=& (\lambda\,x\,c_3+p\,c_1)\,(v-\lambda\,\extd r)\tens v +  (y\,c_3+\lambda\,q\,c_1)\,\extd r\tens v \cr
&&+\,  (\lambda\,x\,c_4+p\,c_2)\,(v-\lambda\,\extd r)\tens \extd r+ (y\,c_4+\lambda\,q\,c_2)\,\extd r\tens \extd r\ .
 \end{eqnarray*}
One can equivalently compute this directly using $i$ and $(\ ,\ )$.  Then $\wedge\,\mathrm{Ricci}=0$ gives the equation
\begin{eqnarray*}
2\,\lambda\,(\lambda\,x\,c_3+p\,c_1)   -  (y\,c_3+\lambda\,q\,c_1)+ (\lambda\,x\,c_4+p\,c_2) \,=\,0\ .
\end{eqnarray*}

Finally, imposing `reality' in the equivalent form on $(\star\tens\id)(\mathrm{Ricci})$ and $\wedge\,\mathrm{Ricci}=0$ gives the 
following values, on the assumption that the denominators do not vanish: 
\begin{eqnarray}\label{nici}
n_1 &=& \frac{b c_3 \lambda ^2+b c_4 \lambda +c_1 \lambda +c_2}{2
   \left(1+b \lambda ^2\right) \left(c_4-\lambda  (c_2-c_3+c_1 \lambda )\right)} \ ,\cr 
   n_2 &=& \frac12 , \cr  n_3 &=&-\frac{1}{2}+\lambda n_1 \cr
   n_4 &=&- \frac{(c_3-\lambda c_1)(c_4+\lambda(c_3-(c_2+\lambda c_1)(2+b\lambda^2)))}{2 (c_1+b
   c_3 \lambda) \left(c_4-\lambda  (c_2-c_3+c_1 \lambda )\right)}.
\end{eqnarray}
Hence $i$ is determined by the symmetry and reality properties of Ricci. We then write the resulting Ricci tensor, as stated.  \endproof

From the Ricci tensor we can of course define the Ricci scalar as before by evaluation with $(\ ,\ )$ to obtain
\begin{eqnarray}\label{Sci} S&:=&(\ ,\ ){\rm Ricci}\nonumber\\
&=&-\frac{(c_2 c_3-c_1 c_4) \left(c_1 \left(1+b \lambda ^2\right)^2+b \left(-c_4+c_2 \lambda  \left(2+b \lambda ^2\right)\right)\right)}{2 r^2 b (c_1+b c_3 \lambda ) (c_4-\lambda  (c_2-c_3+c_1 \lambda ))}.    \end{eqnarray}
We also note that 
\begin{equation}\label{qdim} (\ ,\ )(g)= {2+b\lambda^2\over 1+ b\lambda^2}\end{equation}
plays the role of the `quantum dimension' of our geometry as a kind of trace.

\subsubsection{Example: The decomposable conic family in Proposition~\ref{moduli}}
The conic family (i.e. the decomposable connections according to Corollary~\ref{A1}) has Riemann curvature coefficients computed from (\ref{ciformula}):
\begin{eqnarray*}
c_1&=& \alpha\,(k-1)\cr
c_2&=&  -\frac{4 \alpha ^2 \lambda ^2+b \left(3 \alpha ^2 \lambda ^4-\alpha  \lambda ^2+\alpha 
   k \lambda ^2+k-2\right)}{2 b \lambda } \cr 
   c_3&=&-\frac{4 \alpha ^2 \lambda ^2+b \left(3 \alpha ^2 \lambda ^4+\alpha  \lambda ^2-\alpha 
   k \lambda ^2+k-2\right)}{2 b \lambda } \cr
   c_4&=&  -\frac{\alpha  (k-1) \left(b \lambda ^2+1\right)}{b},
\end{eqnarray*}
where 
\begin{eqnarray*}
k\,=\,\pm \sqrt{4-{\alpha^2 \lambda^2\over b}\left(4 +3 b \lambda^2\right)}\ .
\end{eqnarray*}
Here $+$ corresponds to deformation case of the -ve branch in (\ref{betaroot}). 

This gives lifting map $i$ with
\begin{eqnarray*}
n_1 &=& \frac{4 \alpha ^2 \lambda ^2+b \left(3 \alpha ^2 \lambda ^4-\alpha  \lambda
   ^2+\alpha  k \lambda ^2+k-2\right)}{4 \alpha  (k-1) \lambda  \left(1+b \lambda ^2\right)} \ ,\cr
    n_2&=&\frac12\ ,\quad n_3 =-{1\over 2} + \lambda\, n_1,\quad n_4\,=\,n_1\,{(1+\lambda^2\,b)\over b}
\end{eqnarray*}
and the Ricci tensor
\begin{eqnarray}\label{Ricconic}
\mathrm{Ricci}\ =  \frac{2-k}{r^2 (k-1)2\alpha \lambda^2}  g\ 
\end{eqnarray}
which we find is always proportional to the metric.

The zero torsion point $\alpha=8b/(4+7 b\lambda^2)$ (and $k=2(4-b \lambda^2)/(4+7 b\lambda^2)$)  has 
\begin{eqnarray}\label{riccipertn} \mathrm{Ricci}\ = \left( \frac{4+7 b\lambda^2}{ 4-9 b \lambda^2}\right) { g\over r^2}. \end{eqnarray}
By Corollary~\ref{levi}, this is the Ricci curvature of the unique quantum-Levi-Civita connection deforming the classical one for our metric.

We also consider the quantum Einstein tensor and we suppose that this should be defined so as to be conserved with respect to the quantum connection.  If we consider
expressions of the form 
\[ {\rm Einstein}={\rm Ricci}- {S\over \mu} g\]
 then the value of $\mu$ for the entire conic family (\ref{Ricconic}) is determined uniquely by the quantum conservation requirement and necessarily leads to
\[ \mu={2+b\lambda^2\over 1+b\lambda^2},\quad {\rm Einstein}=0.\]
We recognise the value of $\mu$ here as the `quantum diimension' (\ref{qdim}) which means that at least for the conic family
\begin{equation}\label{qeinst} {\rm Einstein}:={\rm Ricci}- {(\ ,\ )({\rm Ricci})\over (\ ,\ )(g)}g.\end{equation}
vanishes. With this definition, we conclude that the noncommutative geometry remains a `vacuum' on the quantum spacetime for the conic part of the moduli space.

\subsubsection{Example: The nonperturbative line of connections  in Proposition~\ref{moduli}}
The family (2) of Proposition~\ref{moduli}, with no classical limit, has Riemann curvature coefficients computed from (\ref{ciformula}) are
\begin{eqnarray*}
c_1&=& -{b(2+b\lambda^2+\delta(1+b\lambda^2))\over (1+b\lambda^2)^2}\cr
  c_2&=& (2-\delta)(2+b\lambda^2)\over \lambda(1+b\lambda^2) \cr
  c_3&=& {-b\lambda^2(3+2b\lambda^2)+\delta(2+ b\lambda^2)(1+b\lambda^2)\over\lambda (1+b\lambda^2)^2} \cr
  c_4&=&{ (2-\delta)(3+2b\lambda^2)\over 1+b\lambda^2}.
  \end{eqnarray*}

The lifting map $i$ comes out as 
\begin{eqnarray*}
n_1 &=& -{(\delta-2)(2+b\lambda^2)\over 2\lambda(2+b\lambda^2+\delta(1+b\lambda^2))}\cr
n_2 &=&{1\over 2} \cr
n_3 &=&-{1\over 2}+\lambda n_1 \cr
n_4 &=&- {(2\delta(1+b\lambda^2)-b\lambda^2)(\delta(3+2 b\lambda^2)-(2+ b\lambda^2))\over 2 b (\delta-2)(2+b\lambda^2+\delta(1+b\lambda^2))}
\end{eqnarray*}
and the Ricci tensor and scalar come out as
\begin{eqnarray}\nonumber
{\rm Ricci}&=&{(\delta-2)\delta(1+b\lambda^2)(4+3 b\lambda^2)\over r^2 2\lambda^2(2+b\lambda^2+\delta(1+b\lambda^2))}(v^*\tens v+\lambda (\extd r\tens v- v^*\tens \extd r)\\ 
&&\quad\quad\quad\quad\quad\quad\quad\quad\quad\quad\quad\quad\quad\quad\quad +{2+3 b\lambda^2-3\delta(1+b\lambda^2)\over b(\delta-2)}\extd r\tens\extd r)\\
S&=& -{\delta (4+3 b\lambda^2)(2\delta(1+b\lambda^2)-b\lambda^2    )\over r^2 2 b\lambda^2(2+b\lambda^2+\delta(1+b\lambda^2))}.
\end{eqnarray}
In this family only the point $\delta=1$, where it intersects with the preceding decomposable family, has nontrivial Ricci proportional to the metric. 

The zero torsion point in this family is at $\delta=(6 + 3 b \lambda ^2)/(2(1+ b \lambda^2)$ and does not particularly simplify. For example, the
 Ricci tensor  and scalar come out as
\begin{eqnarray}\nonumber
{\rm Ricci}&=&-3\frac{(4+3 b\lambda^2)(-2+b\lambda^2)}{20 r^2 \lambda^2 \left(1+b \lambda ^2\right)}(v^*\tens v+\lambda (\extd r\tens v- v^*\tens \extd r)\\ 
&&\quad\quad\quad\quad\quad\quad\quad\quad\quad\quad\quad\quad\quad + \frac{\left(1+b \lambda ^2\right) \left(14+3 b \lambda ^2\right)}{ b \left(-2+b \lambda ^2\right)}   \extd r\tens\extd r)\\
S&=&-\frac{3 \left(3+b \lambda ^2\right) \left(4+3 b \lambda ^2\right)}{5 r^2  b \lambda ^2 \left(1+b \lambda ^2\right)}.
\end{eqnarray}
After some computation, we find that there is no linear combination of the form ${\rm Ricci}- {S\over \mu} g$ that is conserved with respect to the quantum covariant derivative, which should not shock us as the automatic vanishing of the Einstein tensor in 2D is a classical phenomenon and while it may deform to perturbative solutions in the quantum case as we have seen above, there is no reason to think it also holds in the `deep quantum' regime. 

We see also that there is one  zero Ricci curvature point  in this family namely  $\delta=0$, with 
\[ c_1=-\frac{b \left(2+b \lambda ^2\right)}{\left(1+b \lambda^2\right)^2},\quad c_2=\frac{4+2 b \lambda^2}{\lambda(1+ +b \lambda ^2)},\quad c_3=-\frac{b \lambda  \left(3+2 b \lambda ^2\right)}{\left(1+b \lambda ^2\right)^2},\quad c_4=\frac{6+4 b \lambda ^2}{1+b \lambda ^2}\]
\[ n_1={1\over\lambda},\quad n_2 ={1\over 2},\quad n_3={1\over 2},\quad n_4={\lambda\over 4}\]
\[ {\rm Ricci}=0,\quad {\rm S=0},\quad T(\extd r)= -{3(2+b\lambda^2)\over 2\lambda(1+b\lambda^2)}{v\wedge\extd r\over r} ,\quad T(v)=\lambda T(\extd r).\]
This could be viewed as a nonperturbative vacuum solution of Einstein's equations but with torsion. 

We also have the non-classical situation  in dimension 2 of  a zero Ricci-scalar point where the {\rm Ricci} tensor itself does not vanish, namely $\delta=b\lambda^2/(2(1+b\lambda^2))$. The Riemann coefficients $c_i$ are not very illuminating and we omit them, while
\[ n_1={2+b\lambda^2\over 2\lambda(1+b\lambda^2)},\quad n_2={1\over 2},\quad n_3={1\over 2(1+b\lambda^2)},\quad  n_4=0\]
\[ {\rm Ricci}=-\frac{b \left(4+3 b \lambda ^2\right)}{4(1+ b \lambda ^2)}\Big(v^*\tens v +  \lambda( \extd r\tens v -\, v^*\tens \extd r)-\frac{1+b \lambda ^2}{b} \extd r\tens \extd r\Big)\ ,\quad {\rm S}=0\]
\[  T(\extd r)=- {(3+b\lambda^2)\over \lambda(1+b\lambda^2)}{v\wedge\extd r\over r} ,\quad T(v)=\lambda T(\extd r).\]

\section{Summary and discussion}

In this paper we have fully computed a 2D model of quantum Riemannian geometry based on the simplest possible noncommative spacetime, the 2D version of (\ref{mink}). This is meant as an illustrative model of new phenomena and techniques and we do not draw a very strict physical prediction. However, we believe that the phenomena uncovered are somewhat generic and should also apply in more realistic models. We discuss this here. 

First of all in the 2D model, we showed that the differential algebra (\ref{mink})-(\ref{bicrosscalc}) does not admit a flat quantum metric but only one which is curved, indeed with  singularities in the Ricci tensor along the $t$-axis. This is true even if we then take the classical $\lambda\to 0$ limit. So the idea that spacetime is quantum has predictive power even for {\em classical} GR in the statement that not every classical metric is quantizable. One can analyse this rather more at the strictly semiclassical level and we do this in a sequel \cite{BegMa5}. Briefly, the quantisation is governed by a poisson structure $\omega$ for the quantization of the algebra and a Poisson-compatible linear connection $\nabla$  for the differential structure\cite{BegMa2}. Quantisability of the classical metric $g$ then amounts to a system of equations between $\omega, \nabla, g$ on the classical manifold. They are solved in the 2D bicrossproduct case for the classical metric that we have found in this paper  and not solved by the flat metric. The connection $\nabla$ typically has to be nontrivial to be compatible with $\omega$ and the metric then typically has to be nontrivial to be compatible with $\nabla$. Details will appear in \cite{BegMa5}. 

Our 2D model also had some remarkable behaviour in Section~3.4 for its geodesics, notably the special points $P_\pm$ from which all geodesics emerge or terminate. These are not the `big bang' and `big crunch' in the usual sense, if anything the behaviour of null geodesics resembled more the inside of black hole stretched out to infinity. Nevertheless, if for the sake of crude comparison we identify half the $t$-time,  $1\over\sqrt{-b}$, between these two points with the evolution of our own Universe, we obtain a value for the parameter $-b$,
\[ -b\sim 5\times 10^{-35}\, s^{-2}\ .\]
Note that the $t,r$ coordinate do have a special role both from the original contexts for the algebra (\ref{mink}).  Next, note that in the quantum Riemannian geometry in Section~6 the effective deformation parameter is not $\lambda$ but the dimensionless parameter
\[ b\lambda^2=-b\lambda_P^2\sim 2 \times 10^{-122},\]
see (\ref{qdim}), for example. This gives some idea of the role that we expect the deformation to take. It is around the value of the cosmological constant in natural dimensionless units. 

This may fit in general terms with the proposal in \cite{Ma:newt} for a noncommutative geometry explanation of `vacuum energy' as arising out of an $O(\lambda^2)$ correction from quantum spacetime. In the present context the idea is that Einstein's equation could hold exactly in the quantum geometry but appear not to, i.e. there may be an apparent energy density when the quantum Einstein tensor is expanded as the classical one plus `quantum' corrections. The merit of this approach is that it could explain why the value of the dark energy is so small compared to the natural Planck density, namely because it is a quantum correction one can expect an $O(\lambda^2)$ factor out front. The actual correction depends on the model and on exactly how the quantum algebra and quantum tensors are identified with their classical counterparts, which could take the form of a normal ordering prescription as in \cite{AmeMa}. Thus,  in the 2D model in Section~6 the quantum Einstein tensor, at least as defined, vanishes but if we took the usual Einstein tensor then  at the torsion free `Levi-Civita' point we would have
\[ {\rm Einstein}_{usual}= {\rm Ricci}- {S g\over 2}= \left(1 -{(2+ b\lambda^2)\over 2(1+b\lambda^2)}\right) \left( \frac{4+7 b\lambda^2}{ 4-9 b \lambda^2}\right) { g\over r^2}={b\lambda^2\over 2 r^2}g+O(\lambda^3).\]
We see that there is an $O(b\lambda^2)$ correction to the classical Einstein tensor that is proportional to $g/r^2$.  The metric $g$ itself contains an $O(\lambda)$ correction compared to the classical but this contributes at higher order and we may as well use the classical metric here. Such a correction could be viewed as a non-constant `dark energy' cosmological term. Such non-constant terms are not  conserved but nevertheless could  have a dynamic or `interacting vacuum' cosmological interpretation\cite{DWW}.  There may, however,  be other corrections to the classical Einstein tensor coming from the identification of the tensors (this requires a fuller semiclassical analysis). 

In the $n>2$  case we showed the differential algebra (\ref{mink})-(\ref{bicrosscalc}) does not admit any central metric at all meaning that we need to work with a slightly more general formalism. We do not need the quantum metric $g\in \Omega^1\tens_A\Omega^1$ to be central in order to have a well-defined notion of quantum metric-compatible and quantum torsion (hence of `Levi-Civita') connections and their quantum Riemannian curvature -- at this level the formalism already exists. The problem concerns the inverse metric $(\ ,\ )$ and contractions made with it as explained in the introduction, and this is needed for example in the Ricci tensor and the notion of divergence. Non-centrality will then lead to certain effects  which could, however, be contained if the metric is at least central with regard to some physically relevant subalgebra $A_0\subseteq A$. At the semiclassical level centrality of the quantum metric corresponds to the Poisson-compatible connection $\nabla$ being metric compatible\cite{BegMa5},  cf\cite{BegMa2} for centrality of the symplectic structure and this was the root of the constraint that we have encountered. It is, however, possible to consider metrics  preserved only in some directions and to develop a theory at this reduced level. 

This was the line taken in the present paper where we focussed on a spherically symmetric setting as should be of interest in cosmology.  We again obtained a constraint on the form of the classical metric for it to be quantisable in some weaker sense where only centrality with functions of $r,t$ was required. After that the story at the classical level was not too different from the 2D case, with a 2-parameter moduli of classical metrics all with curvature decaying as $1\over r^2$ and  in some cases an Einstein tensor that implies a perfect fluid for suitable pressure and density as in Section~3.3. The origin of such a fluid would still need to be explained. One of the cases could fit into a quintessence model with $w_Q=-{1\over 2}$ but that model itself would need to be found. The other case with density $\rho=0$ is even more unusual. These could be interesting directions for further work, as would development of the general formalism based on a pair $A_0\subseteq A$. 

Assuming we stick to (\ref{mink}), another get-out within the existing formalism could be to change the differential structure (\ref{bicrosscalc}) to one that is not translation-invariant or, even more extreme, to one that is not associative. The first does not require any change to the formalism and it may be that there is a calculus for which there exists a central metric, just not translation-invariant. Freeing up the calculus amounts to choosing $\nabla$ more freely. The nonassociative route is also possible, even with the Minkowski metric\cite{Beg}, although full details of the formalism of quantum Riemannian geometry in the nonassociative case would need to be developed along the lines of \cite{BegMa3}. At the semiclassical level it is known\cite{BegMa2} that nonassociativity corresponds to the Poisson-compatible connection $\nabla$ having curvature which seems reasonable enough. The main problem with both non-translation invariance and curvature is that once we allow these there are too many possibilities for $\nabla$; we need a physically motivated field equation to further constrain the functional degrees of freedom.  This is another direction suggested by our results. 

\appendix
\section{Christoffel symbols for the classical metric}

Unlike the noncommutative geometry computations, the ones for the Christoffel symbols for the classical metric in Section~3 are routine and hence  relegated to this appendix. The symbols are easily computed from (\ref{gij}) and the formula (this is recalled to fix conventions),
\begin{eqnarray*}
    \Gamma_{jk}^l = \frac{1}{2}\sum_r g^{lr} (\partial _j g_{rk} + \partial _k g_{jr} - \partial _r g_{jk} ).
\end{eqnarray*}
Then
\begin{eqnarray*}
 \Gamma_{jk}^4 = \frac{1}{2\,r^2\,\sin^2\theta}(\partial _j g_{4k} + \partial _k g_{j4}  )
\end{eqnarray*}
so that all  $\Gamma_{jk}^4$ are zero, except for $\Gamma_{4k}^4=\Gamma_{k4}^4$, given by
\begin{eqnarray*}
\Gamma_{42}^4 \ =\ \frac{1}{r}\ ,\quad \Gamma_{43}^4 \ =\ \cot\theta\ .
\end{eqnarray*}
Similarly
\begin{eqnarray*}
    \Gamma_{jk}^3 = \frac{1}{2\,r^2}(\partial _j g_{3k} + \partial _k g_{j3} - \partial _\theta g_{jk} )
\end{eqnarray*}
so that all $\Gamma_{jk}^3$ are zero, except for 
\begin{eqnarray*}
\Gamma_{13}^3\ =\ \Gamma_{31}^3 \ =\ \frac{1}{r}\ ,\quad 
\Gamma_{44}^3\ =\ -\sin\theta\, \cos\theta\ .
\end{eqnarray*}

Similarly, we need to compute
\begin{eqnarray*}
   2\, \Gamma_{jk}^1 &=& g^{11} (\partial _j g_{1k} + \partial _k g_{j1} - \partial _t g_{jk} ) +g^{12} (\partial _j g_{2k} + \partial _k g_{j2} - \partial _r g_{jk} ) \cr
   2\, \Gamma_{jk}^2 &=& g^{21} (\partial _j g_{1k} + \partial _k g_{j1} - \partial _t g_{jk} ) +g^{22} (\partial _j g_{2k} + \partial _k g_{j2} - \partial _r g_{jk} ) 
\end{eqnarray*}
Now, putting $k=4$ gives
\begin{eqnarray*}
   2\, \Gamma_{j4}^1 &=& g^{12} ( - \partial _r g_{j4} ) \cr
   2\, \Gamma_{j4}^2 &=&  g^{22} ( - \partial _r g_{j4} )\ .
\end{eqnarray*}

Now, putting $k=3$ gives
\begin{eqnarray*}
   2\, \Gamma_{j3}^1 &=& g^{11} (\partial _\theta g_{j1} - \partial _t g_{j3} ) +g^{12} ( \partial _\theta g_{j2} - \partial _r g_{j3} ) \cr
   2\, \Gamma_{j3}^2 &=& g^{21} (\partial _\theta g_{j1} - \partial _t g_{j3} ) +g^{22} ( \partial _\theta g_{j2} - \partial _r g_{j3} )\ .
\end{eqnarray*}
Now neither $g_{j1}$ nor $g_{j2}$ depends on $\theta$, and $g_{j3}$ does not depend on $t$, so
\begin{eqnarray*}
   2\, \Gamma_{j3}^1 &=& g^{12} (  - \partial _r g_{j3} ) \cr
   2\, \Gamma_{j3}^2 &=& g^{22} ( - \partial _r g_{j3} )\ .
\end{eqnarray*}
Now we only have $\Gamma_{jk}^1$ and $\Gamma_{jk}^2$ where $j,k\in\{1,2\}$. Put $k=1$,
\begin{eqnarray*}
   2\, \Gamma_{j1}^1 &=& g^{11} (\partial _j g_{11} + \partial _t g_{j1} - \partial _t g_{j1} ) +g^{12} (\partial _j g_{21} + \partial _t g_{j2} - \partial _r g_{j1} ) \cr
    &=& g^{11} (\partial _j g_{11}  ) +g^{12} (\partial _j g_{21} + \partial _t g_{j2} - \partial _r g_{j1} ) \cr
   2\, \Gamma_{j1}^2 &=& g^{21} (\partial _j g_{11} + \partial _t g_{j1} - \partial _t g_{j1} ) +g^{22} (\partial _j g_{21} + \partial _t g_{j2} - \partial _r g_{j1} ) \cr
   &=& g^{21} (\partial _j g_{11} ) +g^{22} (\partial _j g_{21} + \partial _t g_{j2} - \partial _r g_{j1} )
\end{eqnarray*}
This gives the cases
\begin{eqnarray*}
   2\, \Gamma_{11}^1 
    &=& g^{11} (\partial _t g_{11}  ) +g^{12} (\partial _t g_{21} + \partial _t g_{12} - \partial _r g_{11} ) \cr
        &=& g^{12} (2\,\partial _t g_{21}  - \partial _r g_{11} ) \cr
   2\, \Gamma_{11}^2 
   &=& g^{21} (\partial _t g_{11} ) +g^{22} (\partial _t g_{21} + \partial _t g_{12} - \partial _r g_{11} ) \cr
      &=& g^{22} (2\,\partial _t g_{21}  - \partial _r g_{11} ) \cr
      2\, \Gamma_{21}^1 
    &=& g^{11} (\partial _r g_{11}  ) +g^{12} (\partial _r g_{21} + \partial _t g_{22} - \partial _r g_{21} ) \cr
     &=& g^{11} (\partial _r g_{11}  ) +g^{12} ( \partial _t g_{22}  ) \cr
   2\, \Gamma_{21}^2 
   &=& g^{21} (\partial _r g_{11} ) +g^{22} (\partial _r g_{21} + \partial _t g_{22} - \partial _r g_{21} ) \cr
   &=& g^{21} (\partial _r g_{11} ) +g^{22} ( \partial _t g_{22}  )\ . \cr
\end{eqnarray*}

The last cases are now
\begin{eqnarray*}
   2\, \Gamma_{22}^1 &=& g^{11} (\partial _r g_{12} + \partial _r g_{21} - \partial _t g_{22} ) +g^{12} (\partial _r g_{22} + \partial _r g_{22} - \partial _r g_{22} ) \cr
   &=& g^{11} (2\,\partial _r g_{12}  - \partial _t g_{22} ) +g^{12} (\partial _r g_{22}  ) \cr
   2\, \Gamma_{22}^2 &=& g^{21} (\partial _r g_{12} + \partial _r g_{21} - \partial _t g_{22} ) +g^{22} (\partial _r g_{22} + \partial _r g_{22} - \partial _r g_{22} )   \cr
   &=& g^{21} (2\,\partial _r g_{12}  - \partial _t g_{22} ) +g^{22} (\partial _r g_{22}  )\ .
\end{eqnarray*}

We are now ready to obtain all the following Christoffel symbols $\Gamma^k_{ij}$, written as matrices with row $i$ and column $j$,
\begin{eqnarray}\label{christoffel}
\Gamma^1_{\bullet\bullet} &=& \left(
\begin{array}{cccc}
 -\frac{2 b t}{a} & \frac{a+2 b t^2}{a r} & 0 & 0 \\
 \frac{a+2 b t^2}{a r} & -\frac{2 t \left(a+b t^2\right)}{a r^2} & 0 & 0
   \\
 0 & 0 & -\frac{t}{a} & 0 \\
 0 & 0 & 0 & -\frac{t \sin ^2(\theta)}{a}
\end{array}
\right) \cr
\Gamma^2_{\bullet\bullet} &=& \left(
\begin{array}{cccc}
 -\frac{2 b r}{a} & \frac{2 b t}{a} & 0 & 0 \\
 \frac{2 b t}{a} & -\frac{2 b t^2}{a r} & 0 & 0 \\
 0 & 0 & -\frac{r}{a} & 0 \\
 0 & 0 & 0 & -\frac{r \sin ^2(\theta)}{a}
\end{array}
\right)  \cr
\Gamma^3_{\bullet\bullet} &=& \left(
\begin{array}{cccc}
 0 & 0 & 0 & 0 \\
 0 & 0 & \frac{1}{r} & 0 \\
 0 & \frac{1}{r} & 0 & 0 \\
 0 & 0 & 0 & -\sin (\theta) \cos (\theta)
\end{array}
\right) \cr
\Gamma^4_{\bullet\bullet} &=& \left(
\begin{array}{cccc}
 0 & 0 & 0 & 0 \\
 0 & 0 & 0 & \frac{1}{r} \\
 0 & 0 & 0 & \cot (\theta) \\
 0 & \frac{1}{r} & \cot (\theta) & 0
\end{array}
\right)\ .
\end{eqnarray}

\end{document}